\documentclass[aps,pra,showpacs,onecolumn]{revtex4-1}
\usepackage{amssymb}
\usepackage{amsmath}
\usepackage{graphicx}
\usepackage{epsfig}

\begin{document}

\title{Sudden death of particle-pair Bloch oscillation and unidirectioanl
propagation in a one-dimensional driven optical lattice}
\author{S. Lin${}^1$, X. Z. Zhang${}^2$ and Z. Song${}^1$}
\email{songtc@nankai.edu.cn}
\affiliation{${\ }^1$School of Physics, Nankai University, Tianjin 300071, China \\
${\ }^2$College of Physics and Materials Science, Tianjin Normal University, Tianjin 300387, China}


\begin{abstract}
We study the dynamics of bound pairs in the extended Hubbard model driven by
a linear external field. It is shown that the two interacting bosons or
singlet fermions with nonzero on-site and nearest-neighbor interaction
strengths can always form bound pairs in the absence of the external field.
There are two bands of bound pair, one of which may have incomplete wave
vectors when it has a overlap with the scattering band, referred as
imperfect band. In the presence of the external field, the dynamics of the
bound pair in the perfect band exhibits distinct Bloch-Zener oscillation
(BZO), while in the imperfect band the oscillation presents a sudden death.
The pair becomes uncorrelated after the sudden death and the BZO never comes
back. Such dynamical behaviors are robust even for the weak coupling regime
and thus can be used to characterize the phase diagram of the bound states.
\end{abstract}
\pacs{03.65.Ge, 05.30.Jp, 03.65.Nk, 03.67.-a}
\maketitle


\section{Introduction}

The dynamics of particle-pair in lattice systems has received a lot of
interest, due to the rapid development of experiment. Ultra-cold atoms have
turned out to be an ideal playground for testing few-particle fundamental
physics since optical lattices provide clean realizations of a range of
many-body Hamiltonians. It stimulates many experimental \cite{Winkler,Folling,Gustavsson}\ and theoretical investigations in strongly correlated systems, which mainly focus on the
bound-pair formation \cite{Mahajan,Valiente2,MVExtendB,MVTB,Javanainen,Wang}, detection \cite{Kuklov}, dynamics \cite{Petrosyan,Zollner,ChenS,Valiente1,Wang,JLNJP,Corrielli},
collision between single particle and pair \cite{JLBP,JLTrans}, and bound-pair condensate \cite{Rosch}.
The essential physics of the proposed bound
pair (BP) is that, the periodic potential suppresses the single particle
tunneling across the barrier, a process that would lead to a decay of the
pair. This situation cannot be changed in general case when a weak linear
potential is applied. Then a BP acts as a single particle, sharing the
single-particle dynamical features, such as Bloch oscillation (BO),
Bloch-Zener oscillation (BZO) \cite{MUGA, Poladian, Greenberg, Kulishov,
Ruschhaupt, Longhi2012}.

The aim of this paper is to show that the nearest-neighbor (NN) interaction
can not only lead to distinct BO and BZO, but also induce the sudden death
of the oscillations within a Bloch period.\ We study the dynamics of BPs
 in the extended Hubbard model driven by a linear external field. It is
shown that two interacting bosons or singlet fermions with nonzero on-site
and nearest-neighbor interaction strengths can always form BPs in
the absence of the external field. There are two kinds of BP, which
forms two bound bands. We find that the existence of the nearest-neighbor
interaction can lead to the overlap between the scattering band of a single
particle and the bound band, which can spoil the completeness of the bound
band, referred as imperfect band. In the presence of the external field, the
dynamics of the BP in the perfect band exhibits perfect BO and BZO,
while in the imperfect band\ the oscillation presents a sudden death. The
pair becomes uncorrelated after the sudden death of the oscillation and the
correlation never come back. This behavior is of interest in both
fundamental and application aspects. It can be utilized to control the
unidirectional propagation of the BP wavepacket by imposing a single
pulse, which is of great interest for applications in cold atom physics.
Numerical simulations are shown that this scheme achieves very high efficiency
and wide spectral band. Such a unidirectional filter for cold-atom pair may
be realized in a shaking optical lattice in experiment.

This paper is organized as follows. In Section \ref{sec_model}, we present
the model Hamiltonian, and the two-particle band structures. In Section \ref%
{sec_BP dynamics}, we investigate the BP dynamics in the
presence of linear field. Section \ref{sec_Unidirectional propagation} is
devoted to the application of our finding, the realization of unidirectional
propagation induced by a pulsed field. Finally, we give a summary and
discussion in Section \ref{sec_Summary}.

\section{Model Hamiltonian and band structure}

\label{sec_model}
\begin{figure}[tbp]
\centering
\includegraphics[ bb=18 7 575 527, width=0.49\textwidth, clip]{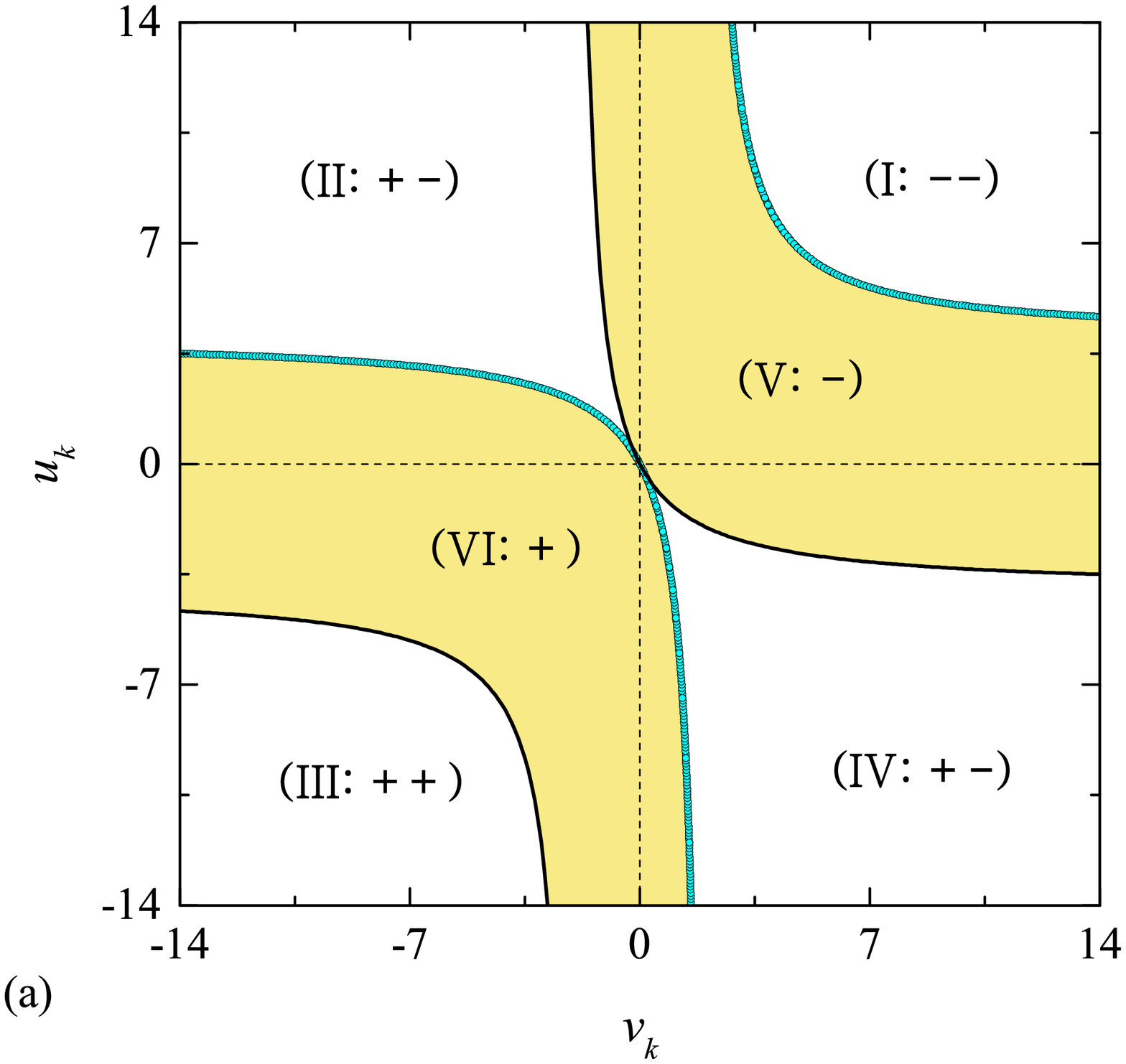} %
\includegraphics[ bb=18 7 575 527, width=0.49\textwidth, clip]{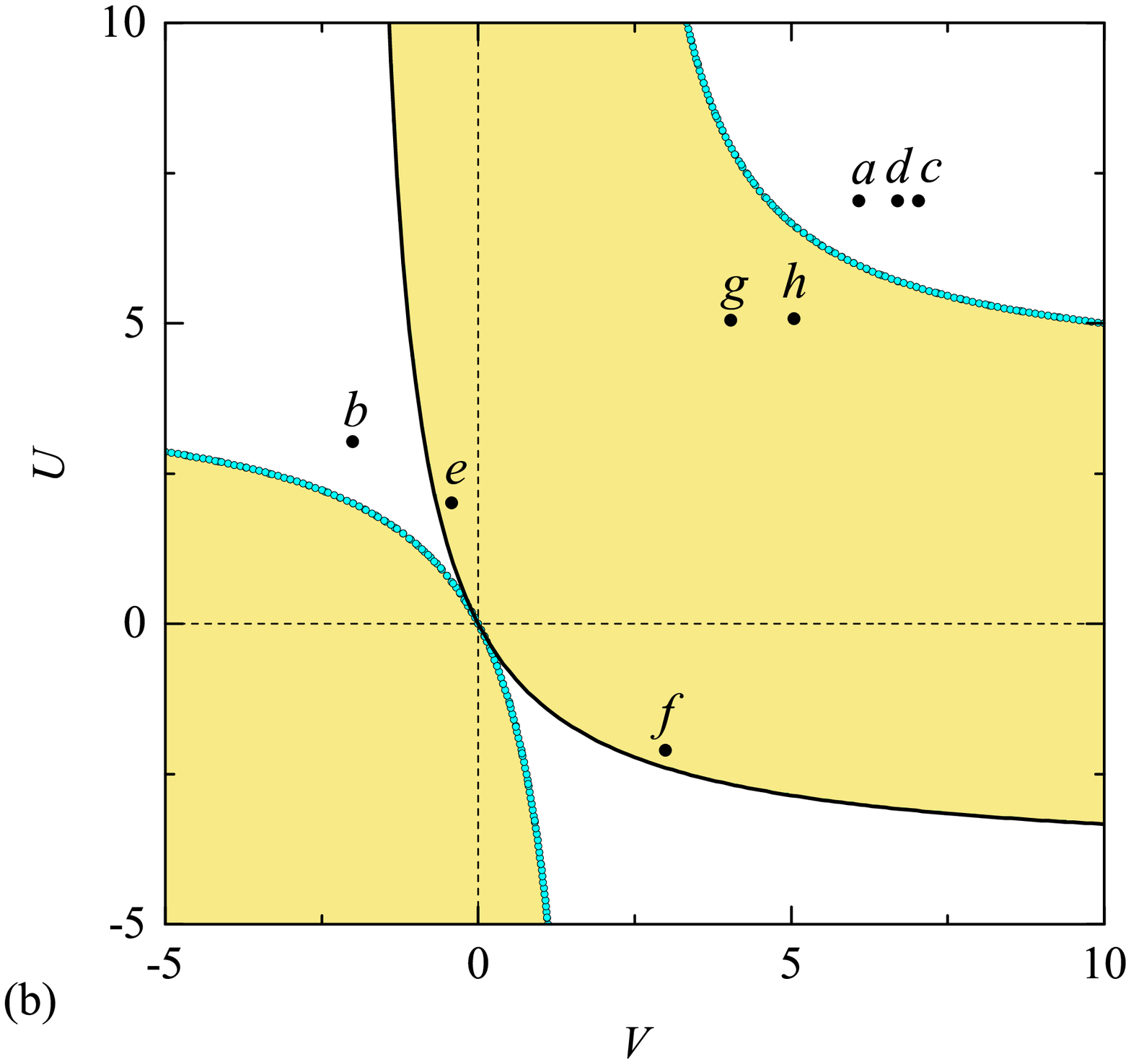}
\par
\caption{(Color online) (a) Phase diagram for the BP states. The
solid black (blue) line is the hyperbolic function in Eq. (\protect\ref%
{boundary}), which is the boundary between the transition from the BP state $\left\vert \protect\psi _{k}^{+}\right\rangle $ ($\left\vert
\protect\psi _{k}^{-}\right\rangle $) to a scattering state in each
invariant subspace indexed by $k$. There are six regions divided by the
lines from Eq. (\protect\ref{boundary}), where $\pm $ indicates the types of
bound states $\left\vert \protect\psi _{k}^{\pm }\right\rangle $ in each
regions.\ (b) Phase diagram (in unit of $\protect\kappa $) indicates the
feature of the BP band: complete band in regions \textrm{I-IV},
while incomplete band in regions \textrm{V} and \textrm{VI}. Points $a\left(
7,6\right) $, $b\left( 3,-2\right) $, $c\left( 7,7\right) $, $d\left(
7,6.7\right) $, $e\left( 2,-0.6\right) $, $f\left( -2,3\right) $ , $g\left(
5,4\right) $ and $h\left( 5,5\right) $ denote typical cases in each phases.
The band structures and dynamical features in these cases are presented in
Fig. \protect\ref{fig2} and \protect\ref{fig3}.}
\label{fig1}
\end{figure}

We consider an extended Hubbard model describing interacting particles in
the lowest Bloch band of a one dimensional lattice driven by an external
force, which can be employed to describe ultracold atoms or molecules with
magnetic or electric dipole-dipole interactions in optical lattices. We
focus on the dynamics of the BP states. The pair can be two
identical bosons, or equivalently, spin-$1/2$ fermions in singlet state. For
simplicity we will only consider bosonic systems, but it is straightforward
to extend the conclusion to singlet fermionic pair. We consider the
Hamiltonian
\begin{equation}
H=H_{0}+F\sum_{j=1}jn_{j},  \label{H}
\end{equation}%
where the second term describes the linear external field while $H_{0}$\ is
one-dimensional Hamiltonian for the extended Bose-Hubbard model on a $N$-site
lattice%
\begin{equation}
H_{0}=-\kappa \sum_{j=1}\left( a_{j}^{\dagger }a_{j+1}+\textrm{H.c.}\right) +%
\frac{U}{2}\sum_{j=1}n_{j}\left( n_{j}-1\right) +V\sum_{j=1}n_{j}n_{j+1}.
\label{H_0}
\end{equation}%
where $a_{i}^{\dag }$ is the creation operator of the boson at the $i$th
site, the tunneling strength, on-site and NN interactions between bosons are
denoted by $\kappa $, $U$\ and $V$.

Let us start by analyzing in detail the two-boson problem in this model. As
in Refs. \cite{JLNJP,JLBP,JLTrans}, a state in the two-particle Hilbert
space, can be expanded in the basis set $\left\{ \left\vert \phi
_{r}^{k}\right\rangle,r=0,1,2,...\right\} $, with%
\begin{eqnarray}
\left\vert \phi _{0}^{k}\right\rangle &=&\frac{1}{\sqrt{2N}}\sum_{j} e ^{ %
i kj}\left( a_{j}^{\dag }\right) ^{2}\left\vert \textrm{vac}\right\rangle ,
\\
\left\vert \phi _{r}^{k}\right\rangle &=&\frac{1}{\sqrt{N}}e ^{ i %
kr/2}\sum_{j} e ^{ i kj}a_{j}^{\dag }a_{j+r}^{\dag }\left\vert \textrm{%
vac}\right\rangle ,
\end{eqnarray}%
where $\left\vert \textrm{vac}\right\rangle $\ is the vacuum state for the
boson operator $a_{i}$. Here $k$ denotes the momentum, and $r$ is the
distance between the two particles. Due to the translational symmetry of the
present system, we have the following equivalent Hamiltonian

\begin{eqnarray}
 H_{\mathrm{eq}}^{k}=-J_{k}(\sqrt{2}\left\vert \phi
_{0}^{k}\right\rangle\left\langle \phi _{1}^{k}\right\vert
+\sum_{j=1}\left\vert \phi_{j}^{k}\right\rangle\left\langle \phi
_{j+1}^{k}\right\vert +\textrm{H.c.)}+U\left\vert \phi _{0}^{k}\right\rangle
\left\langle \phi _{0}^{k}\right\vert  
+V\left\vert \phi _{1}^{k}\right\rangle \left\langle \phi _{1}^{k}\right\vert
\label{H_eq}
\end{eqnarray}
in each invariant subspace indexed by $k$. In its present form, $H_{\mathrm{%
eq}}^{k}$ are formally analogous to the tight-binding model describing a
single-particle dynamics in a semi-infinite chain with the $k$-dependent
hopping integral $J_{k}=2\kappa \cos \left( k/2\right) $ in thermodynamic
limit $N\rightarrow \infty $. In this paper, we are interested in the
BP states, which corresponds to the bound state solution of the
single-particle Schr\"{o}dinger equation%
\begin{equation}
H_{\mathrm{eq}}^{k}\left\vert \psi _{k}\right\rangle =\epsilon
_{k}\left\vert \psi _{k}\right\rangle .  \label{S_eq}
\end{equation}%
For a given $J_{k}$, the Hamiltonian $H_{\mathrm{eq}}^{k}$ possesses one or
two bound states, which are denoted as $\left\vert \psi
_{k}^{+}\right\rangle $\ and $\left\vert \psi _{k}^{-}\right\rangle $,
respectively. Here the Bethe-ansatz wavefunctions have the form%
\begin{equation}
\left\vert \psi _{k}^{\pm }\right\rangle =C_{0}^{k}\left\vert \phi
_{0}^{k}\right\rangle +\sum_{r=1}\left( \pm 1\right) ^{r}C_{r}^{k}e^{-\beta
r}\left\vert \phi _{r}^{k}\right\rangle ,
\end{equation}%
with $\beta >0$. For such two bound states $\left\vert \psi _{k}^{\pm
}\right\rangle $ the Schrodinger equation in Eq. (\ref{S_eq}) admits%
\begin{equation}
\pm e^{3\beta }+\left( u_{k}+v_{k}\right) e^{2\beta }\pm \left(
u_{k}v_{k}-1\right) e^{\beta }+v_{k}=0,
\end{equation}%
where $u_{k}=U/J_{k}$ and $v_{k}=V/J_{k}$\ are respectively the reduced
interaction strengthes. The corresponding bound-state energy of $\left\vert
\psi _{k}^{\pm }\right\rangle $ can be expressed as\
\begin{equation}
\epsilon _{k}^{\pm }=\pm J_{k}\cosh \beta .  \label{BS_energy}
\end{equation}%
The transition from bound to scattering states occurs at $\beta =0$. Then
the boundary, at which the bound state $\left\vert \psi _{k}^{\pm
}\right\rangle $ disappears, is described by the hyperbolic function
\begin{equation}
u_{k}=-\frac{2v_{k}}{1\pm v_{k}},  \label{boundary}
\end{equation}%
which is plotted in Fig. \ref{fig1}. It shows that the boundary lines divide
the $u_{k}-v_{k}$ plane into six regions, from $\mathrm{I}$ to $\mathrm{VI}$%
. The type of the bound states in each region can be foreseen from the
extreme situations where $\left\vert u_{k}\right\vert ,$\ $\left\vert
v_{k}\right\vert \gg 1$.\ Under this condition, it is easy to check that
there are two bound states\ with the eigen energies
\begin{equation}
\epsilon _{k}^{\pm }\approx U\textrm{ and }V,  \label{UV}
\end{equation}%
\ in each invariant $k$-subspace. Comparing Eqs. (\ref{UV}) and (\ref%
{BS_energy}), we get the conclusion that there are two bound states in the
regions $\mathrm{I}$,\textrm{\ }$\mathrm{II}$,\textrm{\ }$\mathrm{III}$, and
$\mathrm{IV}$: two $\left\vert \psi _{k}^{-}\right\rangle $ in $\mathrm{I}$,
one $\left\vert \psi _{k}^{+}\right\rangle $ and $\left\vert \psi
_{k}^{-}\right\rangle $\ in $\mathrm{II}$ and $\mathrm{IV}$, two $\left\vert
\psi _{k}^{+}\right\rangle $ in $\mathrm{III}$, while there is a bound state
$\left\vert \psi _{k}^{-}\right\rangle $ in $\mathrm{V}$ and $\left\vert
\psi _{k}^{+}\right\rangle $\ in $\mathrm{VI}$, respectively.

On the other hand, we know that the scattering band of $H_{\mathrm{eq}}^{k}$%
\ ranges from $-4\kappa \cos \left( k/2\right) $ to $4\kappa \cos \left(
k/2\right) $, which reaches the widest bandwidth\ at $k=0$. Therefore, when
we take $J_{0}=\pm 2\kappa $, this diagram can characterize the bound-state
number distribution $N_{\mathrm{b}}(U,V)$: we have $N_{\mathrm{b}}=2N$ in
the regions $\mathrm{I}$,\textrm{\ }$\mathrm{II}$,\textrm{\ }$\mathrm{III}$,
and $\mathrm{IV}$, where all the $2N$ bound states indexed by $k$\
constitute a complete BP band.\ In contrast, we have $N_{\mathrm{b}%
}<2N$ in $\mathrm{V}$ and $\mathrm{VI}$, where all $N_{\mathrm{b}}$ bound
states indexed by survival $k$, which does not cover all range of the
momentum in the Brillouin zone, from $-\pi $ to $\pi $.\ We refer to this
property as incomplete BP band.\ Therefore, the phase diagram also
indicates the boundary $U=-2V/\left( 1\pm V\right) $, for the transition
from the complete to incomplete BP bands, which agrees with the
results reported in Refs. \cite{Valiente2009, Dias, Khomeriki}. For given $U$
and $V$, the complete spectrum of $H_{0}$ can be computed by diagonalizing
the Hamiltonian $H_{\mathrm{eq}}^{k}$ numerically. In Fig. \ref{fig2} and %
\ref{fig3}, we plot the band structures for several typical cases,
which are marked in Fig. 1. We do not cover all the typical points in every
region due to the following fact. The spectrum of $H_{0}$ obeys the relation
\begin{equation}
E_{k}\left( U,V\right) =-E_{k}\left( -U,-V\right) ,
\end{equation}%
in view of%
\begin{equation}
H_{0}(U,V)=-RH_{0}(-U,-V)R^{-1},
\end{equation}%
where the transformation $R$ is defined as $Ra_{j}R^{-1}=\left( -1\right)
^{j}a_{j}$.\textbf{\ }It is a rigorous result in the all range of the
parameters.\textbf{\ }As expected, we\textbf{\ }observe that the
two-particle spectrum comprises\textbf{\ }three Bloch bands, two BP
bands formed by two kinds of BP states, and one scattering band
formed by uncorrelated states.\ We can see from the Fig. \ref{fig2}, that
two bound bands are separated from the scattering band whenever the system
is in the regions (points $a$, $b$, $c$ and $d$) $\mathrm{I}$,\textrm{\ }$%
\mathrm{II}$,\textrm{\ }$\mathrm{III}$, and $\mathrm{IV}$. In contrast,
whenever the points ($e$, $f$, $g$ and $h$) lie in the regions\ $\mathrm{V}$
and $\mathrm{VI}$, the pseudo gap between BP and scattering bands
around $k=0$ vanishes, resulting the formation of incomplete band. What it
quite unexpected and remarkable is that if we apply a linear field, the
dynamics of the BP exhibits some peculiar behaviors, which will be
investigated in the following section.

\section{BP dynamics}

\label{sec_BP dynamics}

Before starting the investigation of the BP dynamics, we would like to study
the relation between the center path of a wavepacket driven by the linear
field and\ dispersion of the Hamiltonian $H_{0}$. Consider a general
one-dimensional tight-binding system, which has the dispersion relation $%
E(k) $ being an arbitrary smooth periodic function $E(2\pi +k)=E(k)$. The
dynamics of wavepacket can be simply understood in terms of the
semiclassical picture: A wavepacket centered around $k_{\mathrm{c}}$ can be
regarded as a classical particle with momentum $k_{\mathrm{c}}$ \cite{Bloch,
Ashcroft, Kittel}. When the wavepacket is subjected to a homogeneous force
of strength $F$, the acceleration theorem $\partial k_{\mathrm{c}}\left(
t\right) /\partial t=F$ tells us
\begin{eqnarray}
k_{\mathrm{c}}\left( t\right) &=&k_{\mathrm{c}}\left( 0\right)
+\int_{0}^{t}Fdt  \nonumber \\
&=&k_{\mathrm{c}}\left( 0\right) +Ft  \label{acceleration theorem}
\end{eqnarray}%
for constant field. The central position of the wavepacket is%
\begin{eqnarray}
x_{\mathrm{c}}\left( t\right) &=&x_{\mathrm{c}}\left( 0\right)
+\int_{0}^{t}\upsilon _{\mathrm{g}}dt  \nonumber \\
&=&x_{\mathrm{c}}\left( 0\right) +\frac{1}{F}\left[ E_{k}\left( k_{\mathrm{c}%
}\left( 0\right) +Ft\right) -E_{k}\left( k_{\mathrm{c}}\left( 0\right)
\right) \right]  \label{CP}
\end{eqnarray}%
where $\upsilon _{\mathrm{g}}=\partial E_{k}/\partial k$ is the group
velocity. Notice that the trajectory of a wavepacket is essentially
identical with the dispersion relation for the field-free system\ under the
semi-classical approximation. This observation provides a fairly clear
picture for the dynamics of a wavepacket in the presence of the linear
field. As a simple example we consider a single-particle case for
illustration. The single-particle BO with Bloch frequency $\omega _{\mathrm{B%
}}=F$ for $H$ can be simply understood from its cosusoidal dispersion
relation $E(k)=-2\kappa \cos k$ rather than quadratic, in momentum $k$.

Now, we switch gears to the case of two-particle. We note that the
bandwidth of the BP band is comparable to that of scattering band,
which leads to the conclusion that a BP wavepacket has distinct
group velocity. It indicates that the dynamics of the BP state has
the similar behavior with the single particle. The BO-like behavior of the
BP wavepacket emerges in the presence of the linear external field.

In order to demonstrate these points, we consider an example for the
Hamiltonian $H$ in Eq. (\ref{H}) with $U$, $V\gg \left\vert U-V\right\vert $%
, $\kappa $. As studied in Ref. \cite{JLNJP}, in the absence of the external
field, the BP lies in the quasi-invariant subspace spanned by the
basis $\{\underline{\left\vert l\right\rangle }\}$, which is defined as%
\begin{equation}
\underline{\left\vert l\right\rangle }\equiv \left\{
\begin{array}{c}
\left( a_{l/2}^{\dag }\right) ^{2}/\sqrt{2}\left\vert \textrm{vac}%
\right\rangle \textrm{, }(\textrm{even } l) \\
a_{\left( l-1\right) /2}^{\dag }a_{\left( l+1\right) /2}^{\dag }\left\vert
\textrm{vac}\right\rangle \textrm{,\ }(\textrm{odd } l)%
\end{array}%
\right. .  \label{BP_basis}
\end{equation}%
In the presence of the external field, the bound-pair can be described by
the following effective Hamiltonian

\begin{equation}
H_{\mathrm{eff}}=-\sqrt{2}\kappa \sum\limits_{l}\left( \underline{\left\vert
l\right\rangle }\underline{\left\langle l+1\right\vert }+\textrm{H.c.}\right)
+\sum\limits_{l}\left[ Fl+\frac{\delta }{2}\left( -1\right) ^{l}\right]
\underline{\left\vert l\right\rangle }\underline{\left\langle l\right\vert },
\label{H_eff}
\end{equation}%
where we neglect a constant term $\left( U+V\right) /2\sum_{l}\underline{%
\left\vert l\right\rangle }\underline{\left\langle l\right\vert }$\ and$\ $%
take $\delta =U-V$ to present the unbalanced on-site and nearest neighbor
interactions. $H_{\mathrm{eff}}$\ is nothing but the tight-binding
Hamiltonian to describe a single particle subjected to a staggered linear
potential, which has been well studied in previous literature \cite{Breid}.
Unlike the fractional BO \cite{Buchleitner, Kolovsky, Kudo2009, Kudo2011} in
the case of $V=0$, $H_{\mathrm{eff}}$ can support wide bandwidth \cite{JLNJP}%
, which is responsible for the large amplitude oscillations. In the
situation with $\delta =0$, it turns out that particle undergoes BO with
frequency $\omega _{\mathrm{B}}=F$. In the case of nonzero unbalance $\delta
\neq 0$, it has been reported that the dynamics of the wavepacket shows a
BZO,\ a coherent superposition of Bloch oscillations and Zener tunnelling
between the sub-bands. The Zener tunnelling takes place almost
exclusively when the momentum of wavepacket reaches $ \pm \pi$. Then we
get the conclusion that the BPs serve as a composite particle,
exhibiting BO and BZO in strong coupling region.

\begin{figure}[tbp]
\centering
\includegraphics[ bb=53 199 532 579, width=0.245\textwidth, clip]{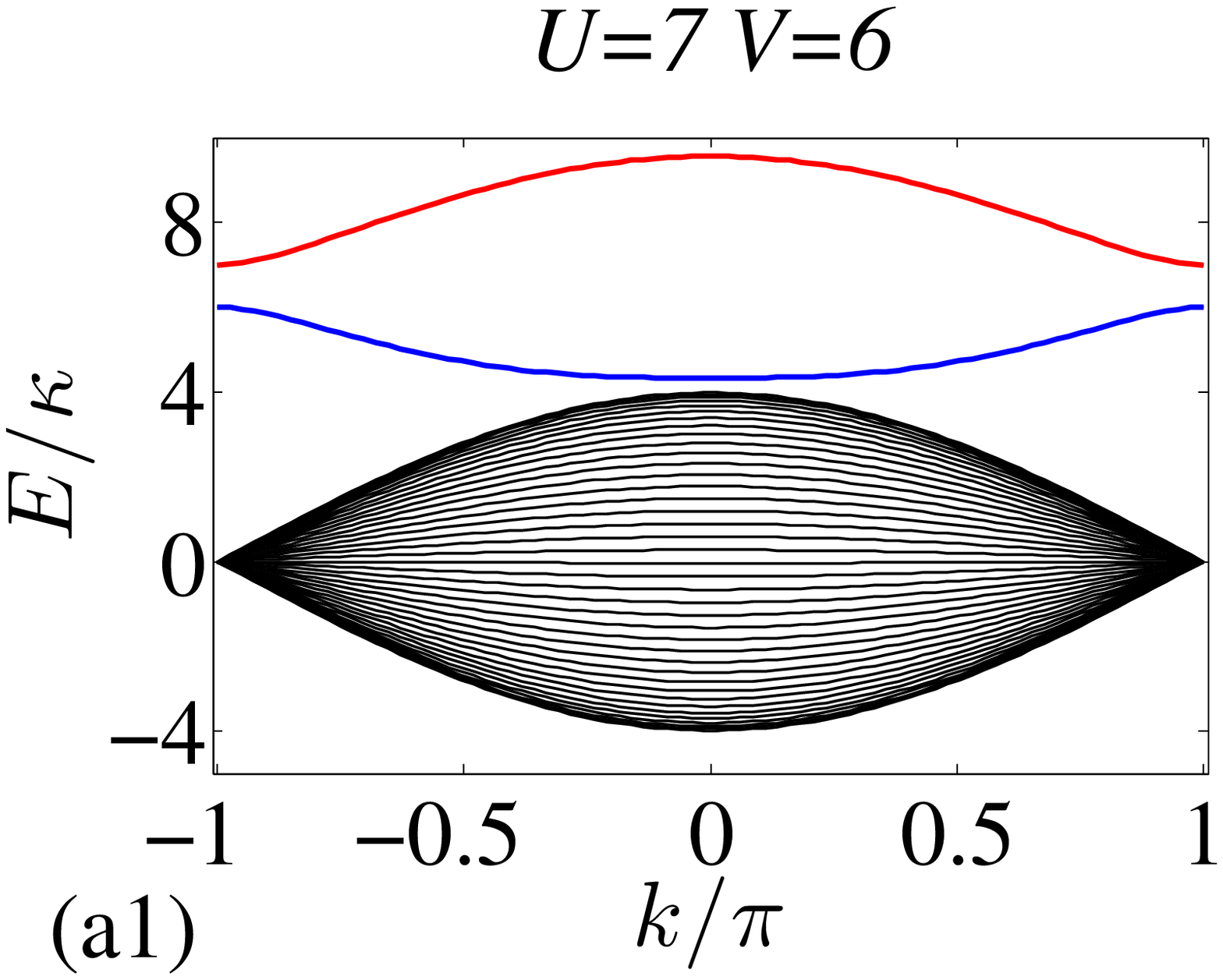} %
\includegraphics[ bb=57 199 532 579, width=0.245\textwidth, clip]{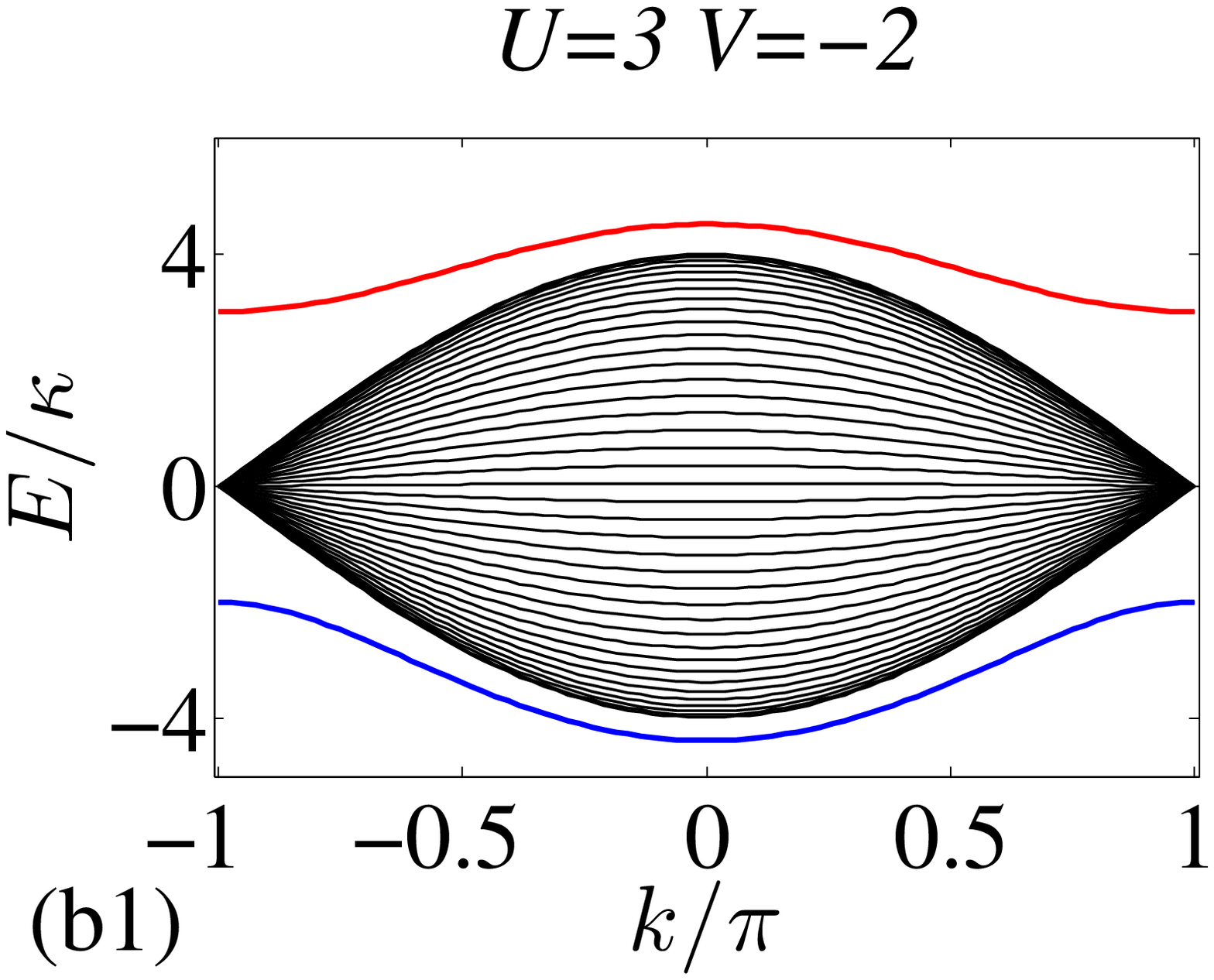} %
\includegraphics[ bb=57 199 532 579, width=0.245\textwidth, clip]{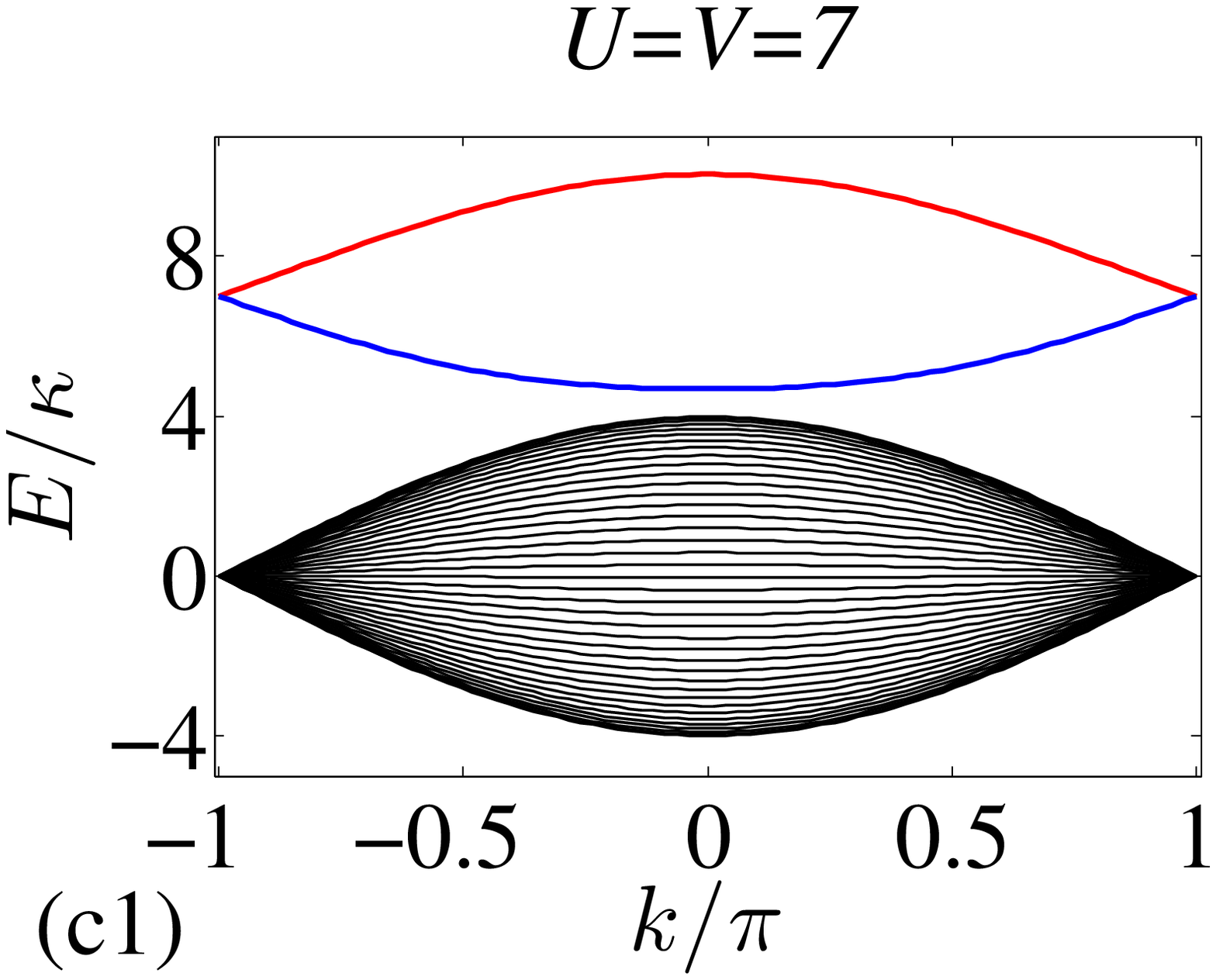} %
\includegraphics[ bb=57 199 532 579, width=0.245\textwidth, clip]{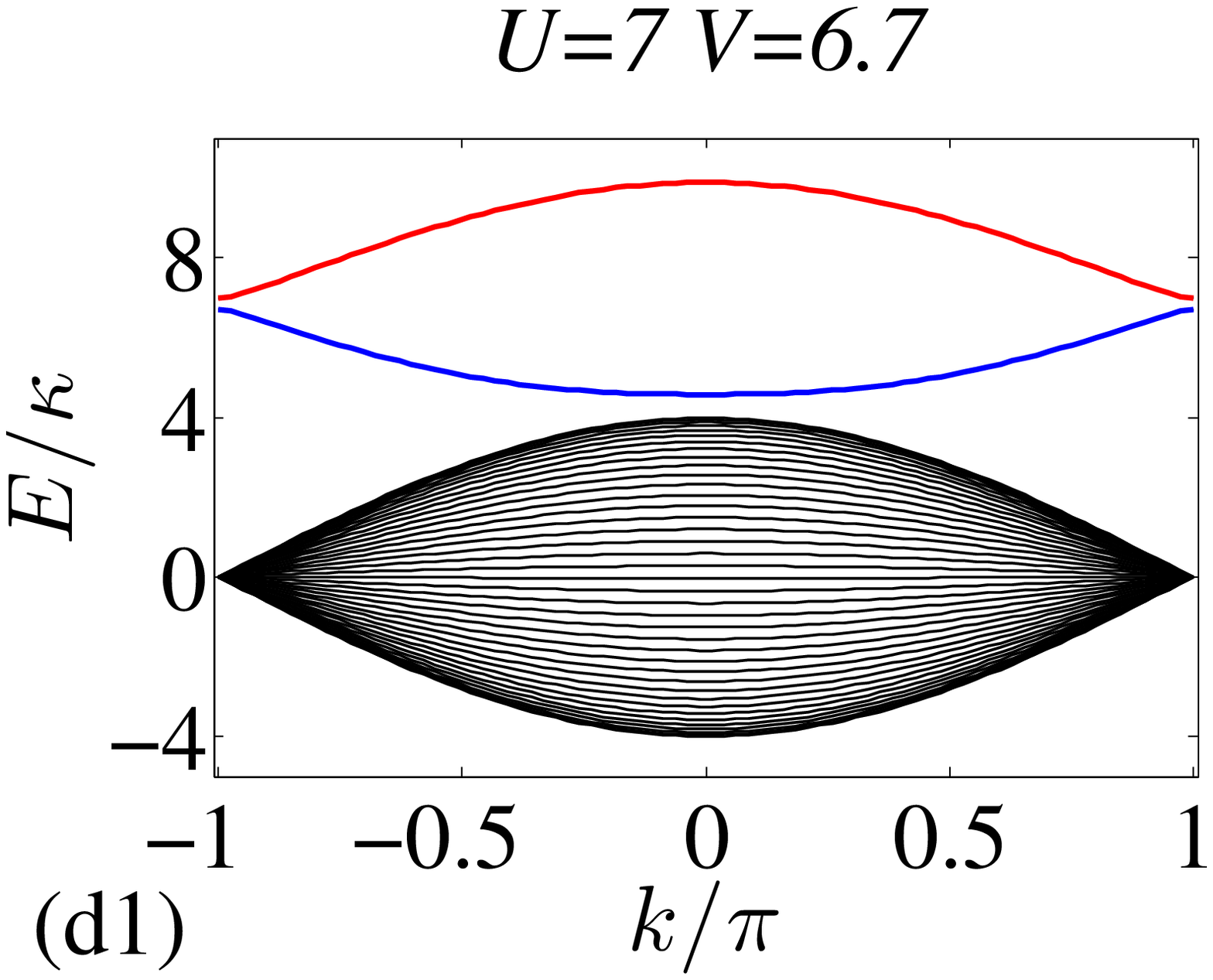} %
\includegraphics[ bb=0 0 635 586, width=0.244\textwidth, clip]{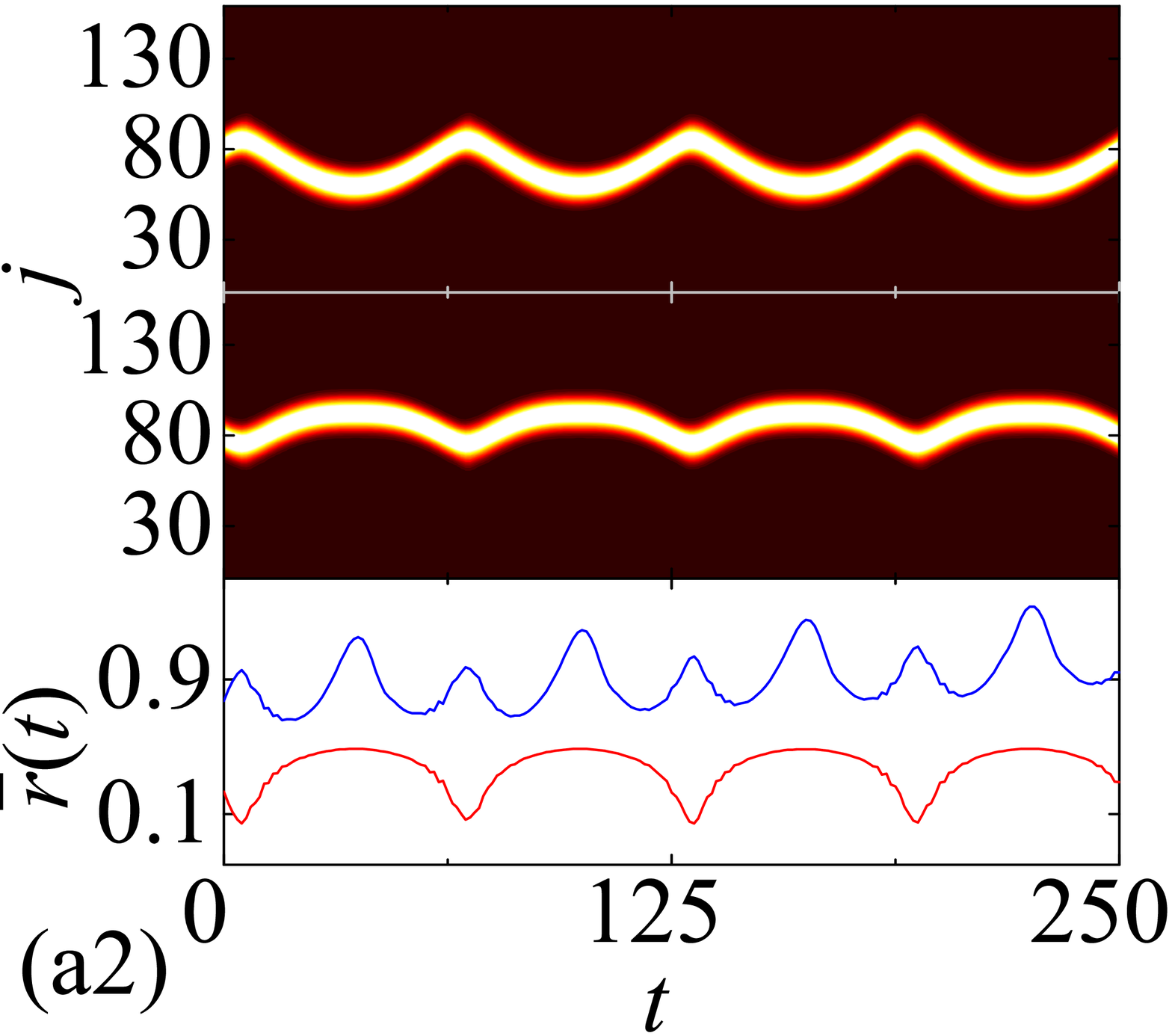} %
\includegraphics[ bb=0 0 635 586, width=0.244\textwidth, clip]{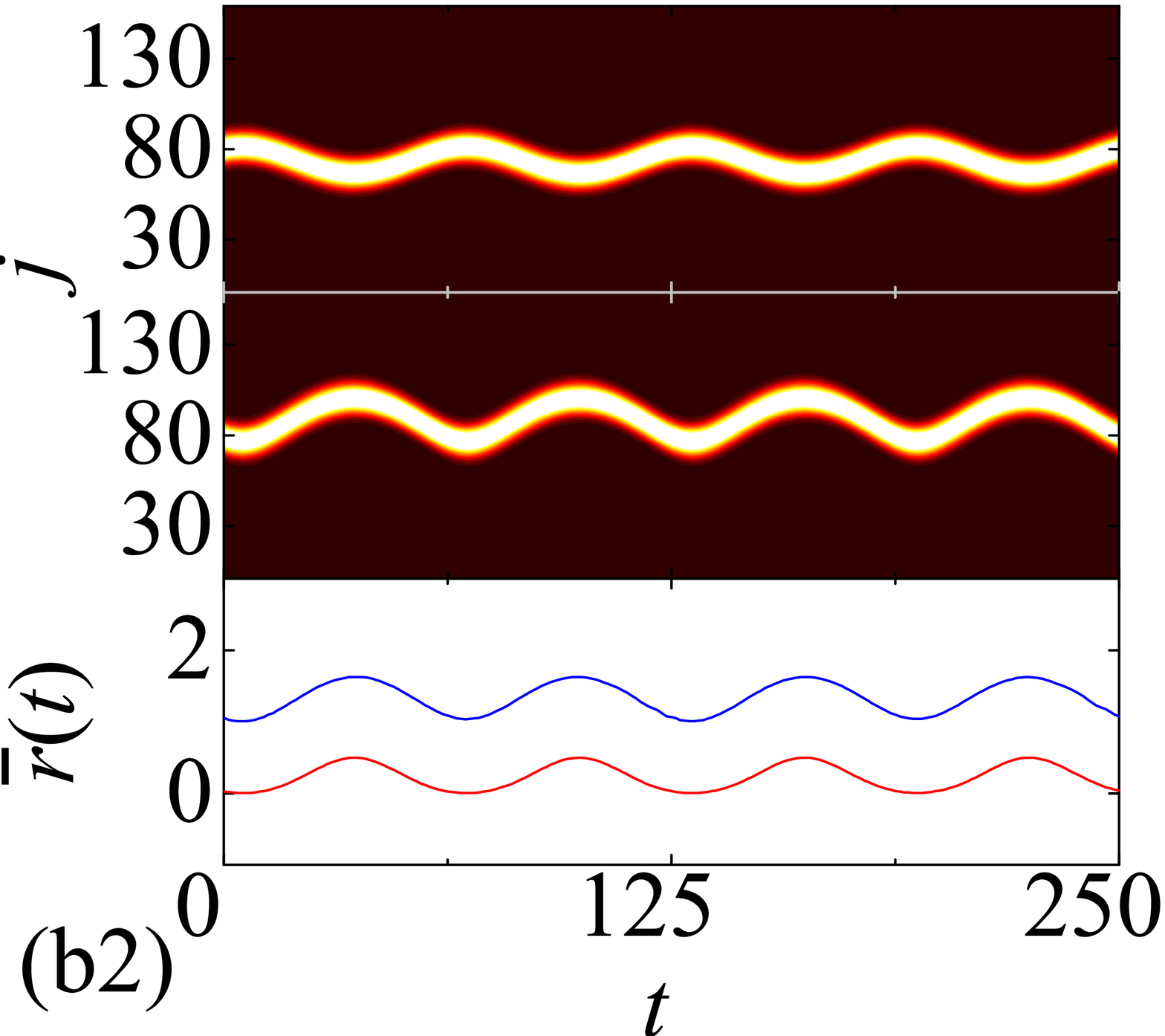} %
\includegraphics[ bb=0 0 635 586, width=0.244\textwidth, clip]{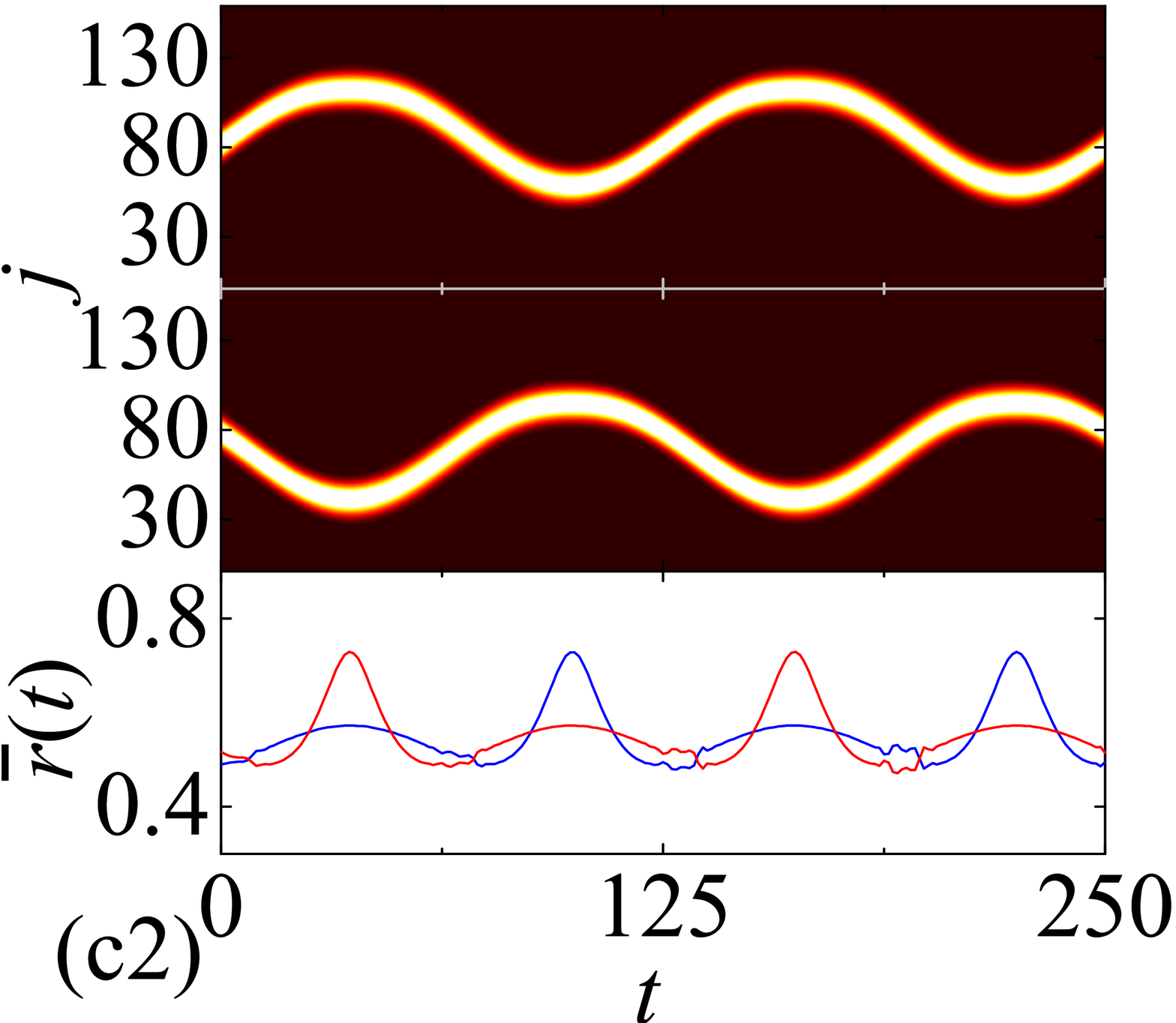} %
\includegraphics[ bb=0 0 635 586, width=0.244\textwidth, clip]{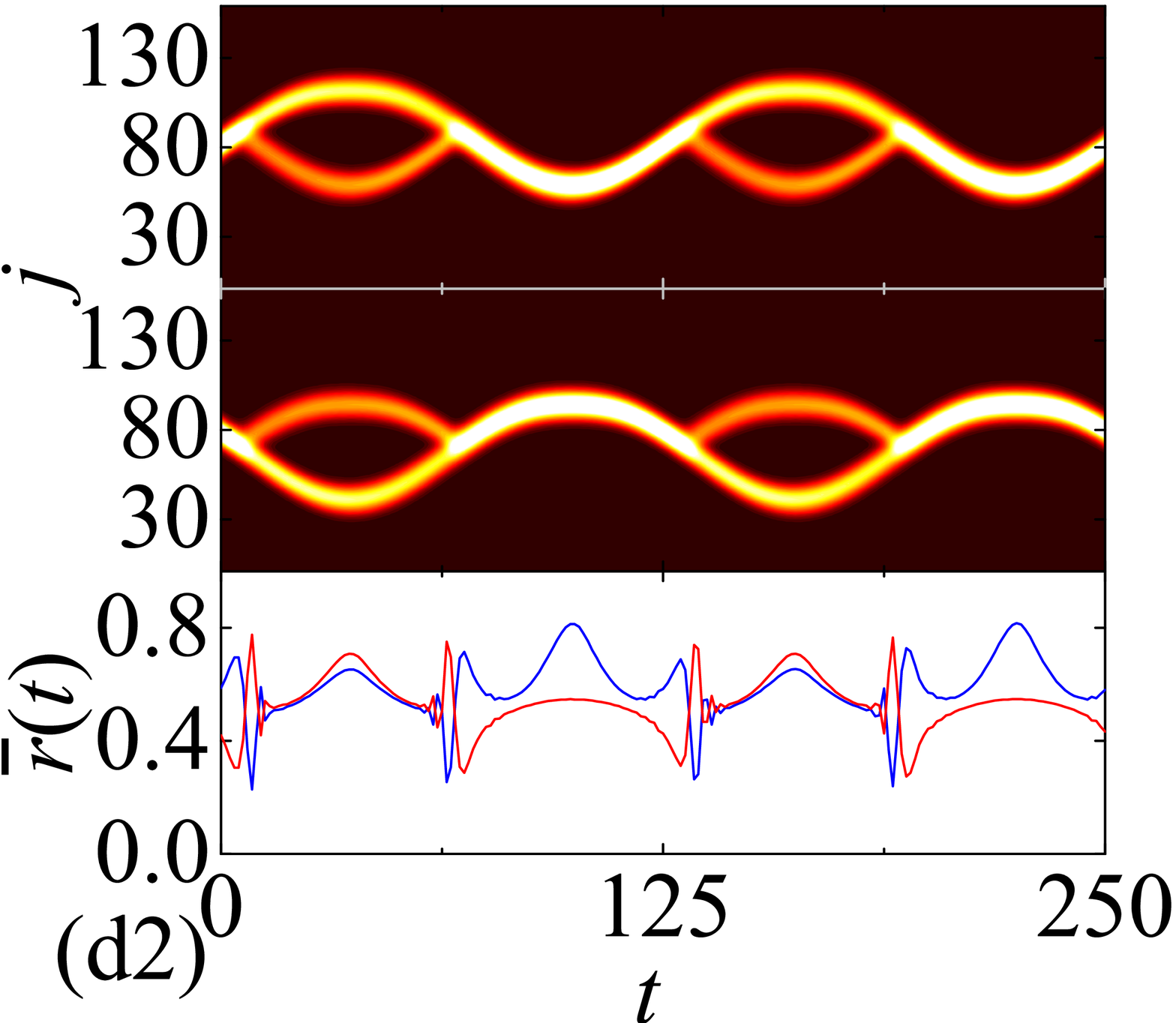}
\par
\caption{(Color online) (a1-d1) The band structures for the systems with
parameters fixed at the typical points ($a$-$d$)\ labelled in Fig. 1. All
the BP bands are complete. (a2-d2) The profiles and the average
distances $\bar{r}\left( t\right) $\ of the time evolution of the initial
wavepackets in the form of Eq. (\protect\ref{Psi_0}) with $k_{0}=-0.8\protect%
\pi $, $\protect\alpha =0.15$, and $N_{\mathrm{A}}=80$, of the two
BP bands, in the external field with $F_{0}=0.05$, in which the time $t$ is expressed in units of $1/\protect\kappa $.\ We can see the
perfect BO and BZO phenomenon. The evolution of $\bar{r}\left( t\right) $
shows that the correlation of the BP remains strong. }
\label{fig2}
\end{figure}

In this paper we are interested in what happens if the initial state is
placed in a incomplete band. It is presumable that the semi-classical theory
still holds when the wavepacket is in the extent of the in incomplete band,
because the nonzero pseudogap can protect the BP wavepacket from the
scattering band. However, when the wavepacket reaches the band edge, the
transition from bound to the scattering band occurs. The wavepacket diffuses
into the continuous spectrum rather than the repetitive motion of
acceleration and Bragg reflection. We refer to this phenomenon as the sudden
death of the BO. In the case of the incomplete BP band with a edge $%
k_{\mathrm{m}}>0$, the life time $\tau $\ for an initial wavepacket with $k_{%
\mathrm{c}}\left( 0\right) $\ satisfies
\begin{equation}
k_{\mathrm{m}}=\left\vert k_{\mathrm{c}}\left( 0\right) +\tau F\right\vert .
\label{tau}
\end{equation}%
When this occurs, the correlation between two particles breaks down and the
wavepacket spreads out in space, irreversibly.

\begin{figure}[tbp]
\centering
\includegraphics[ bb=57 199 532 579, width=0.245\textwidth, clip]{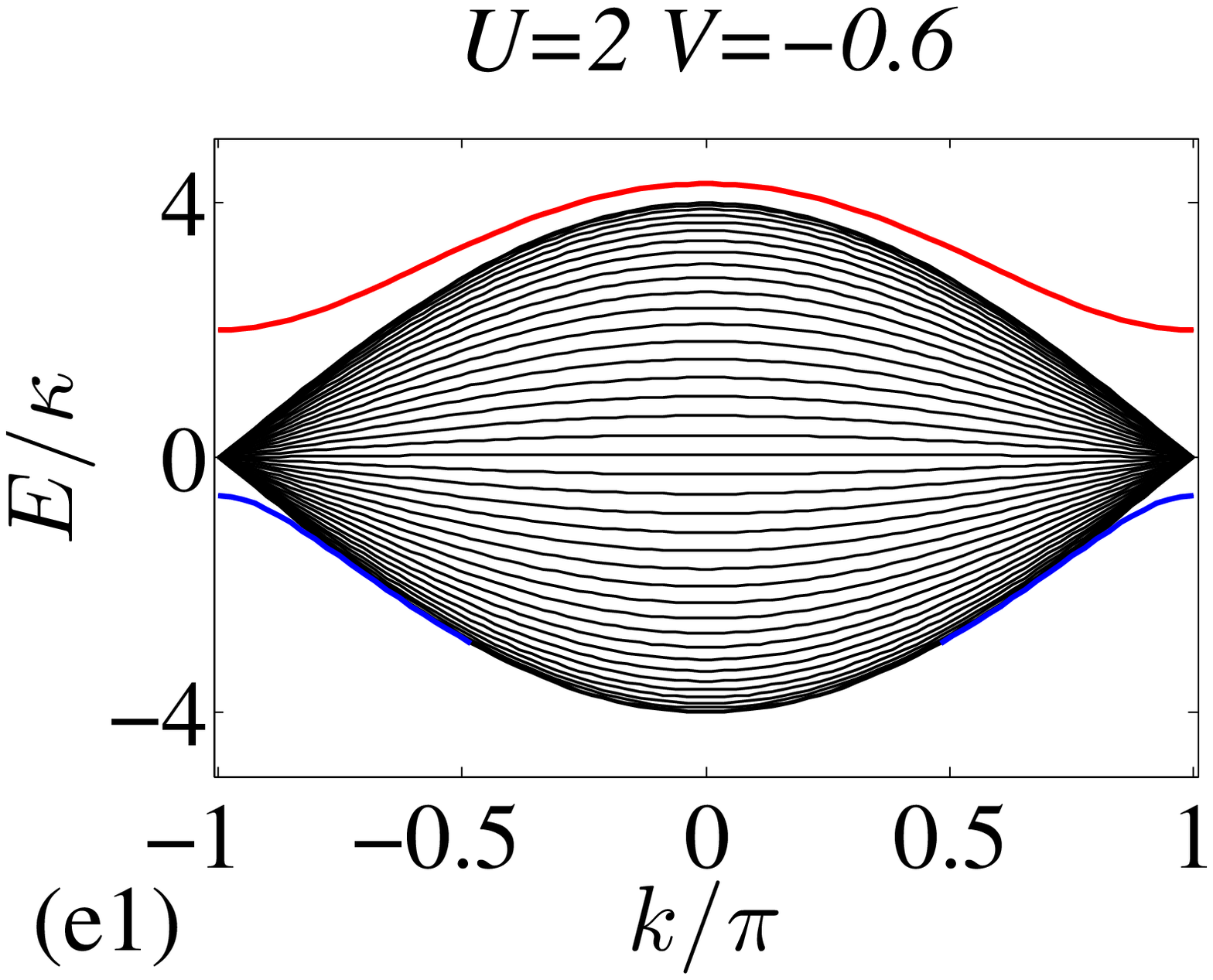} %
\includegraphics[ bb=57 199 532 579, width=0.245\textwidth, clip]{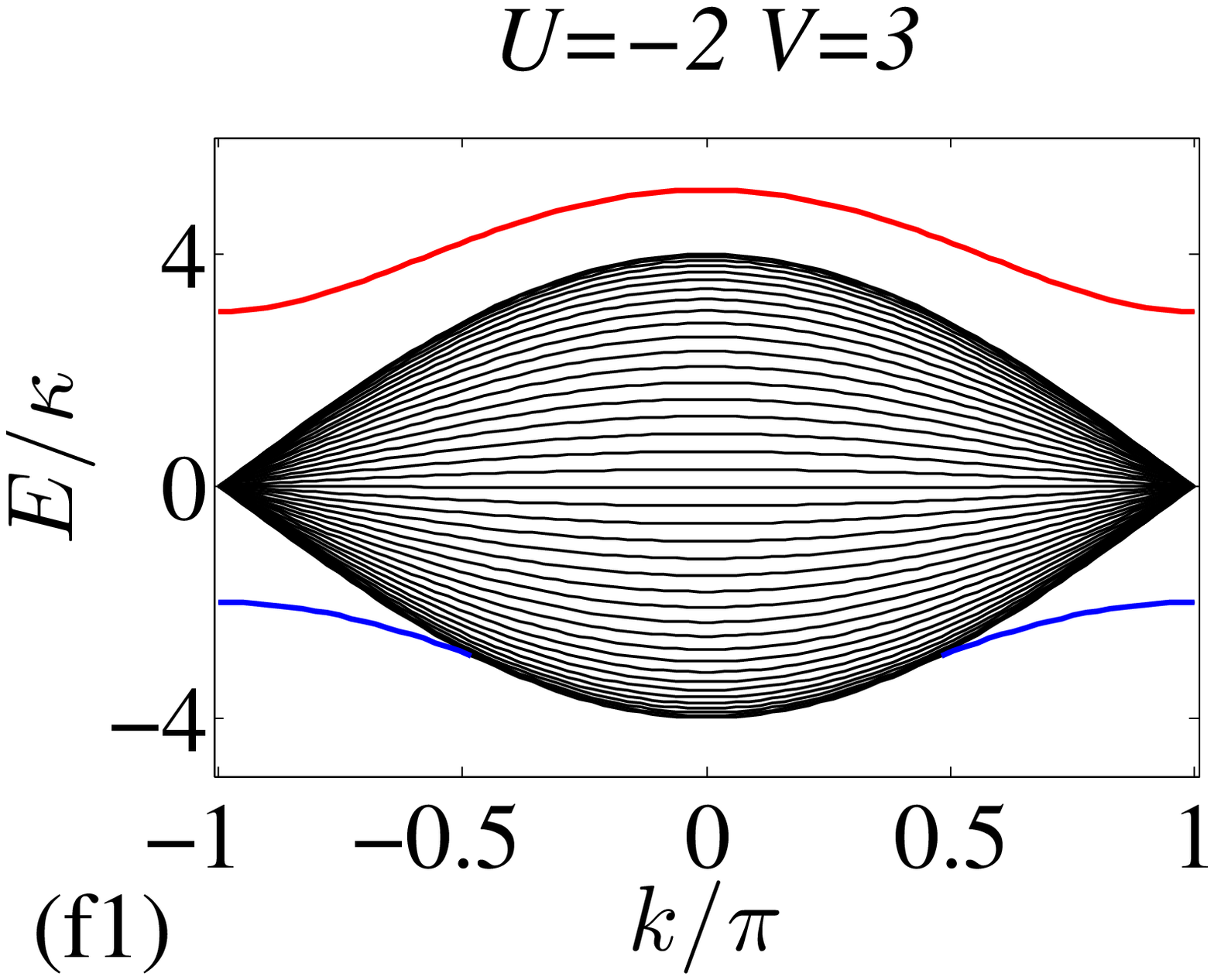} %
\includegraphics[ bb=57 199 532 579, width=0.245\textwidth, clip]{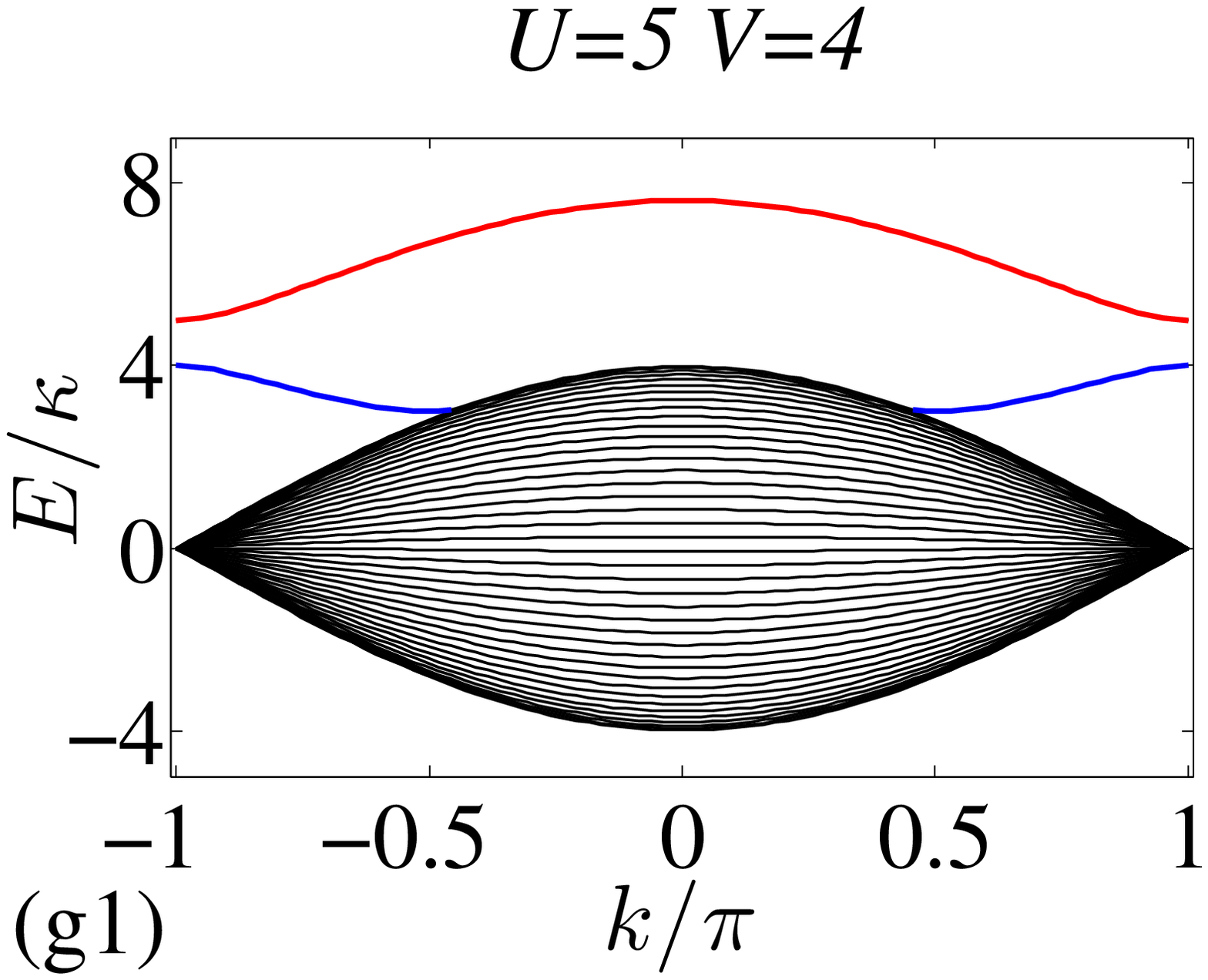} %
\includegraphics[ bb=57 199 532 579, width=0.245\textwidth, clip]{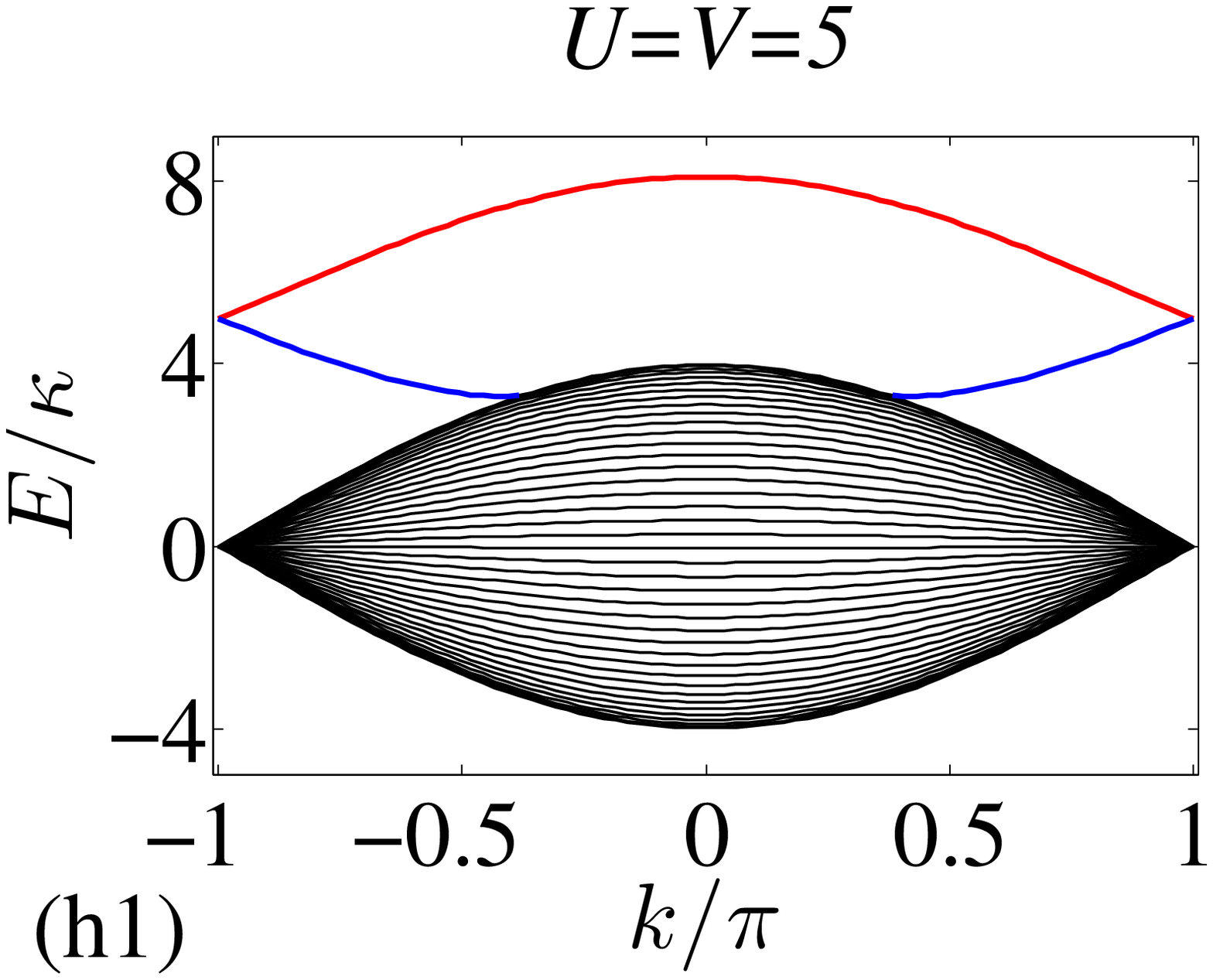} %
\includegraphics[ bb=0 0 635 586, width=0.244\textwidth, clip]{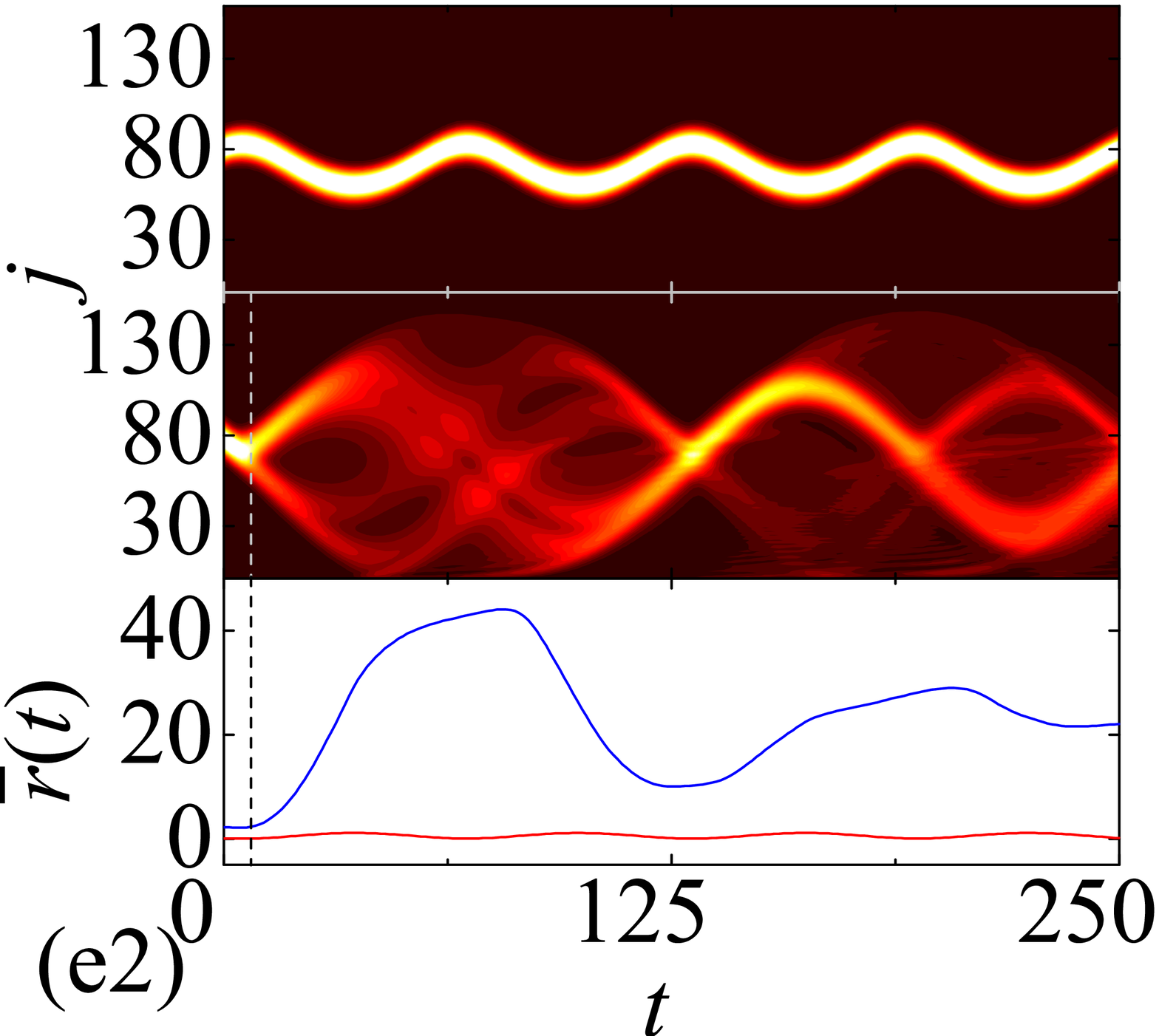} %
\includegraphics[ bb=0 0 635 586, width=0.244\textwidth, clip]{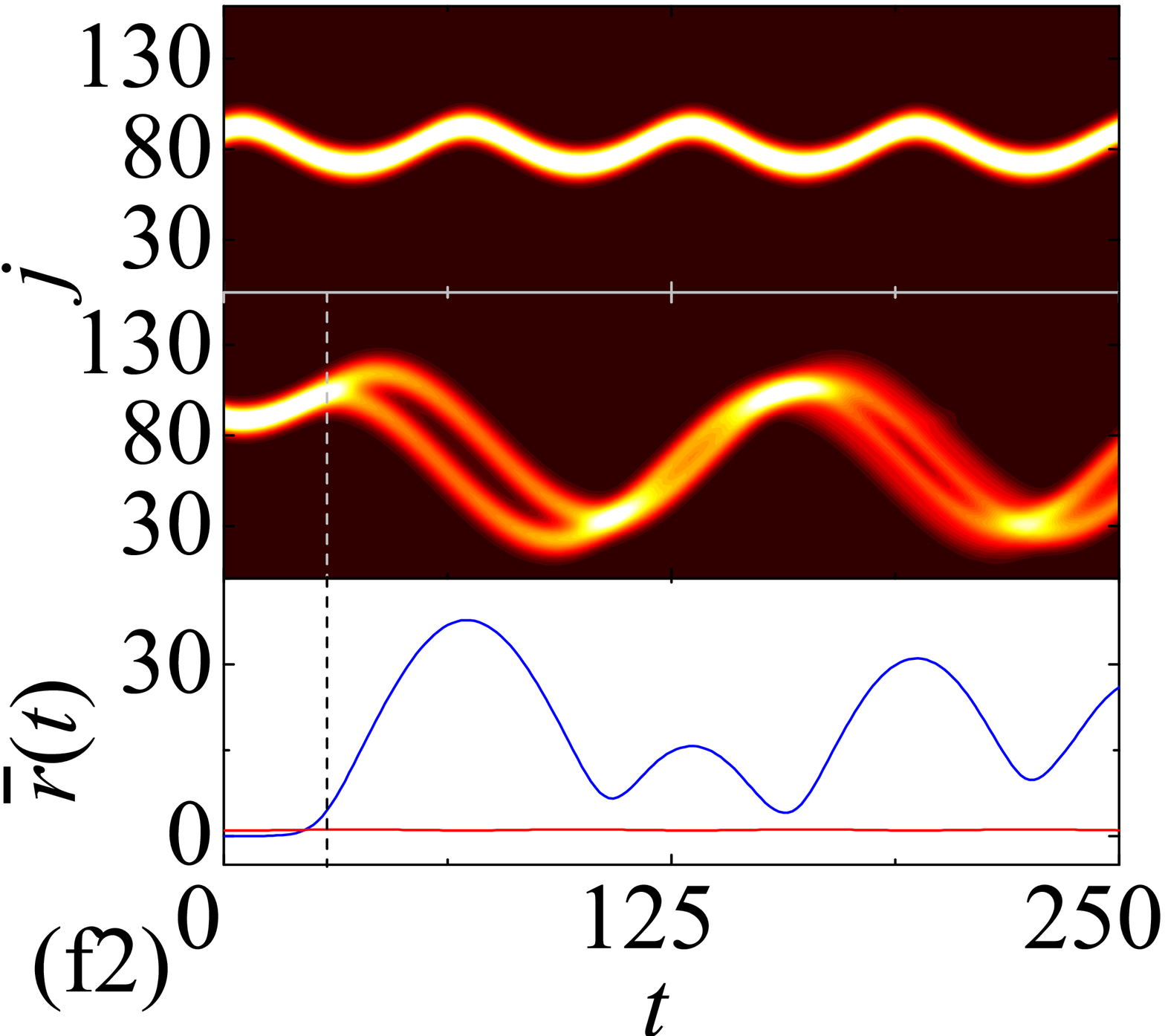} %
\includegraphics[ bb=0 0 635 586, width=0.244\textwidth, clip]{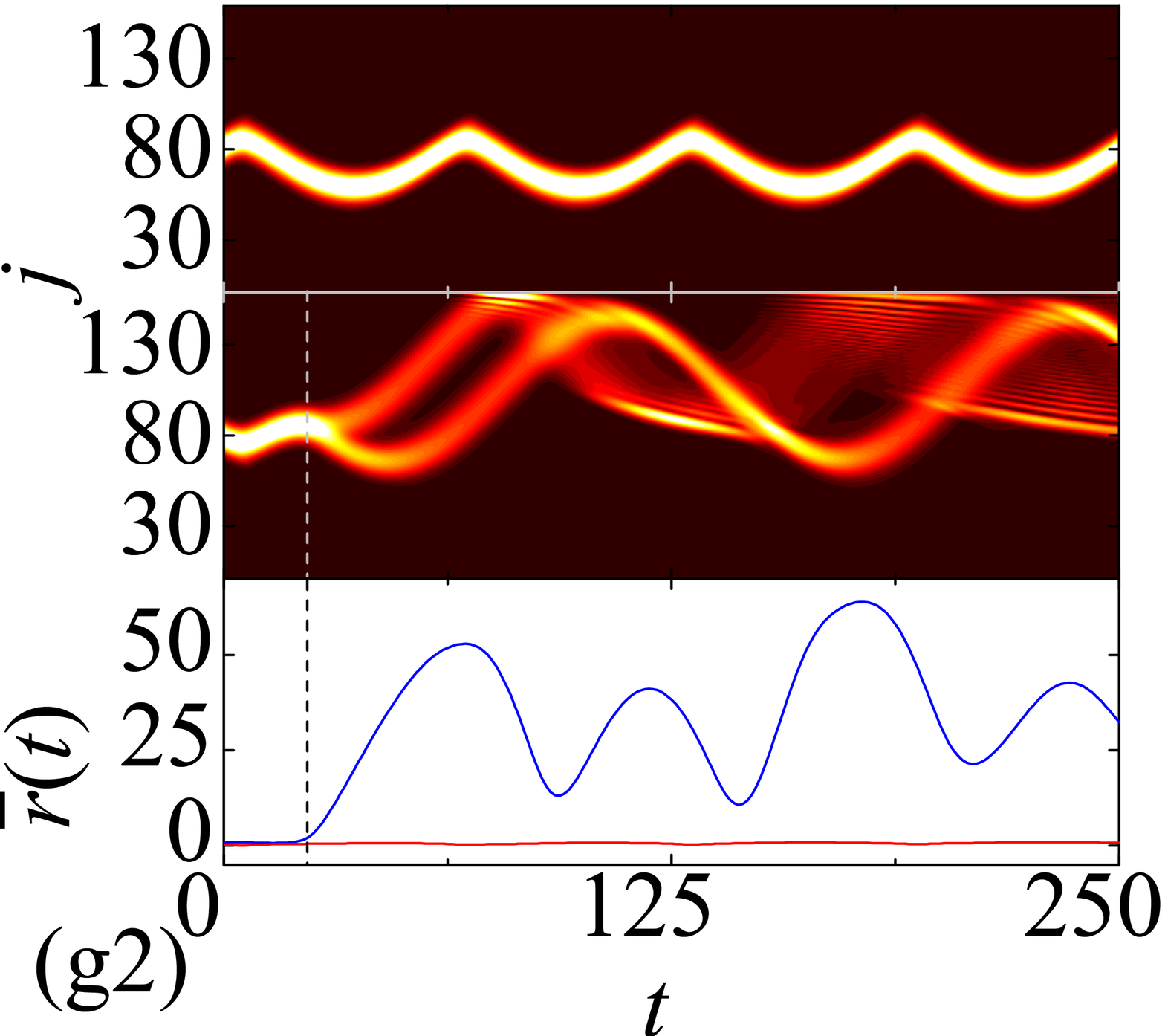} %
\includegraphics[ bb=0 0 635 586, width=0.244\textwidth, clip]{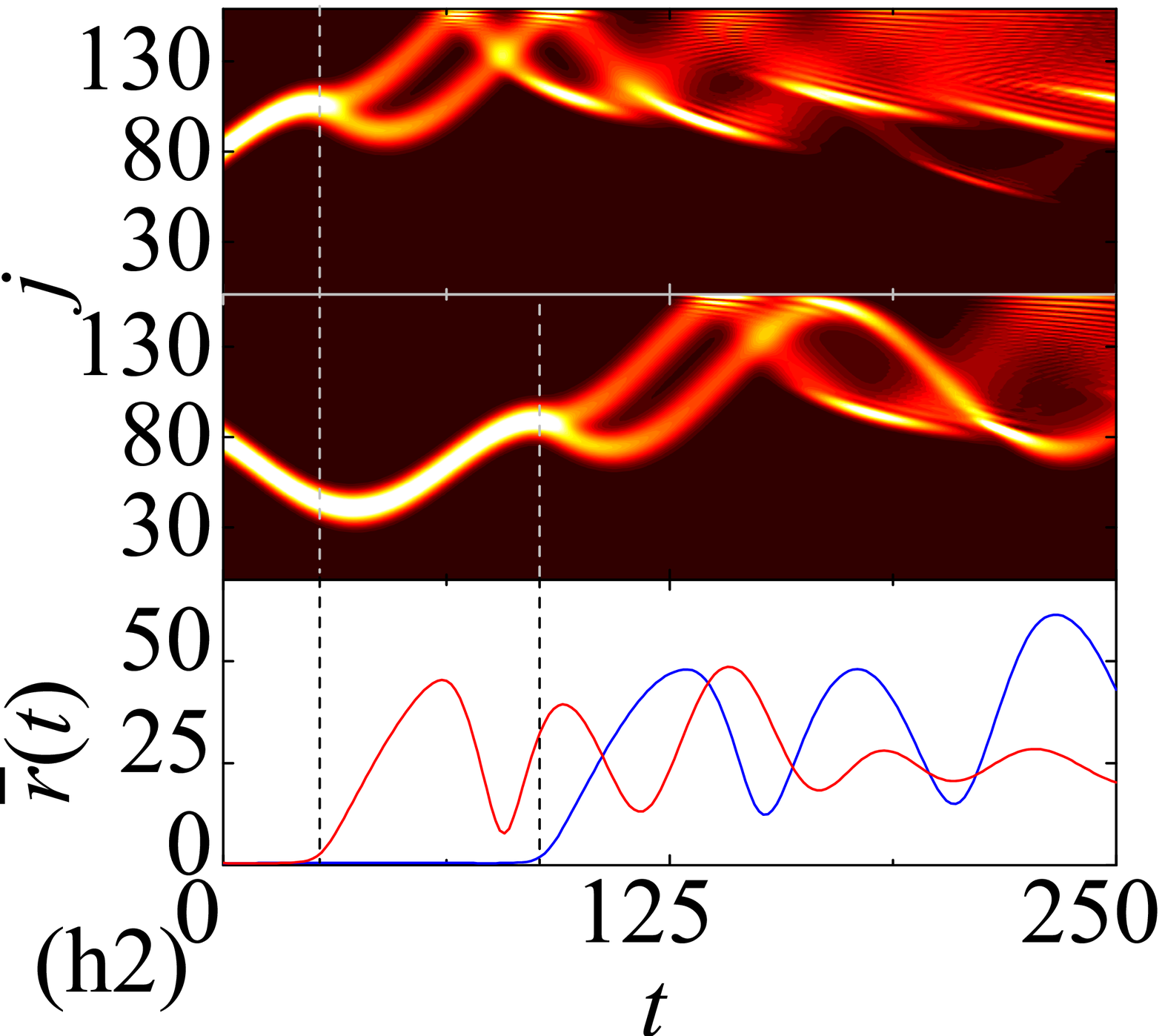}
\par
\caption{(Color online) The same as that in \protect\ref{fig2} but for the
systems with parameters fixed at the typical points $e$-$f$\ labelled in
Fig. 1. It shows that the BP band becomes incomplete due to the
overlap between the bound and scattering energy levels. We can see the
sudden death of the BO, which closely accompanies the breakdown of the pair
correlation. Here the time $t$ is also expressed in units of $1/\protect\kappa $.}
\label{fig3}
\end{figure}

To verify and demonstrate the above analysis, numerical simulations are
performed to investigate the dynamics behavior. We compute the time
evolution of the wavepacket by diagonalizing the Hamiltonian $H$
numerically. Throughout this paper, we investigate the dynamics of the
initial Gaussian wavepacket in the form

\begin{equation}
\left\vert \Psi \left( 0\right) \right\rangle =\Lambda \sum_{k}\exp \left[ -%
\frac{\left( k-k_{0}\right) ^{2}}{2\alpha ^{2}}-i N_{\mathrm{A}}\left( k-k_{0}\right)\right]
\left\vert \psi _{k}\right\rangle , \label{Psi_0}
\end{equation}%
where $\Lambda $ is the normalization factor, $k_{0}$ and $N_{\mathrm{A}}$
denote the central momentum and position of the initial wavepacket,
respectively. The evolved state under the Hamiltonian $H$ is $\left\vert \Psi \left( t\right) \right\rangle$ = $ e^{-i Ht}\left\vert\Psi \left( 0\right) \right\rangle  \nonumber$.
We would like to stress that the initial wavepacket involves solely one
BP band, either upper or lower one. However, the evolved state may
involve two BP bands, even the scattering band,\ when the Zener
tunnelling occurs. We plot the probability profile of the wavepacket
evolution in several typical cases in Fig. \ref{fig2} and \ref{fig3}. In
Fig. \ref{fig2}, the simulation is performed in the systems, where two bound
bands are well separated from the scattering band. As the external field is
turned on, several dynamical behaviors occur: when two bound bands are well
separated ($a1$ and $b1$), the BOs in both bands are observed ($a2$ and $b2$%
). In the case ($d1$, $d2$), two bound bands are very close at $\pm \pi $,
which induces the BZO as expected. For the three cases, the BO
frequency doubles comparing with that of single particle case. Case ($c1$,
$c2$)\ fixes $U=V$, two bound bands merg into a single bound band. As
predicted above, simple\ BO rather\ than\ BZO is\ observed, with a period
equal to that of single-particle BO.
In Fig. \ref{fig3}, the systems have a common feature: one of the BP
bands is incomplete due to the pseudo gap vanishing. In the three cases ($e2$%
, $f2$ and $g2$), the BOs remain in one band, whereas the BOs break down at
the edges of the incomplete band. For the case ($h1$, $h2$) with $U=V$,\ two
bound bands merg into a single incomplete band. The BP wavepackets
in both two bands cannot survive from the irreversible spreading.

Furthermore, the correlation between two particles is measured by the
average distance of two particle
\begin{equation}
\overline{r}\left( t\right) =\sum_{i,r}r\left\langle \Psi \left( t\right)
\right\vert n_{i}n_{i+r}\left\vert \Psi \left( t\right) \right\rangle ,
\end{equation}%
which can be used to characterize the feature of sudden death of BO for a
evolved state $\left\vert \Psi \left( t\right) \right\rangle $. As
comparison, the average distance $\overline{r}\left( t\right) $ as function
of time for several typical cases is plotted in Fig. \ref{fig2} and \ref%
{fig3}. We find that the sudden death of BO is always accompanied by the
irreversible increasing of $\overline{r}\left( t\right) $, which accords
with our analytical predictions.

\begin{figure}[tbp]
\centering
\includegraphics[ bb=0 0 594 594, width=0.32\textwidth, clip]{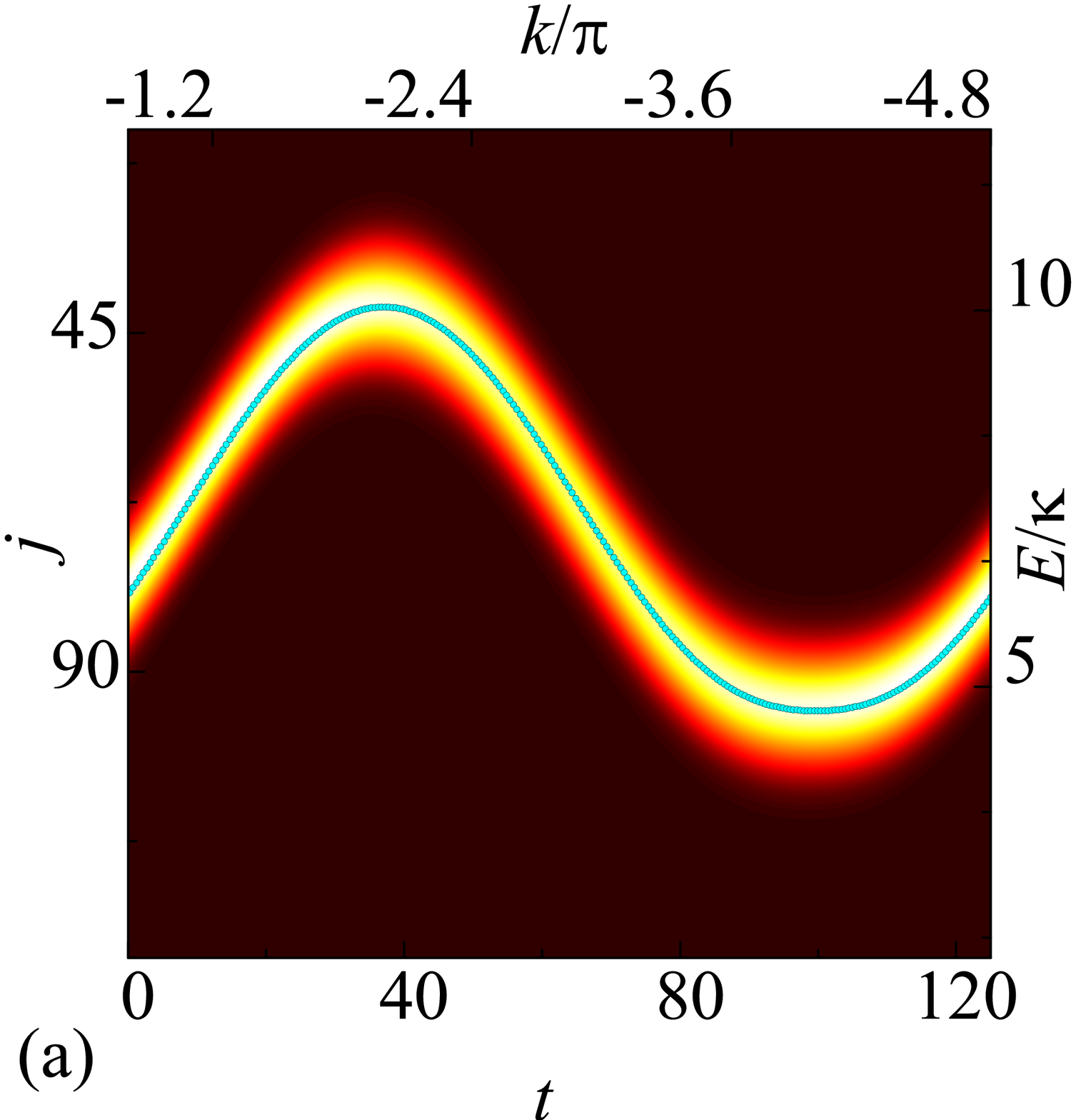} %
\includegraphics[ bb=0 0 594 594, width=0.32\textwidth, clip]{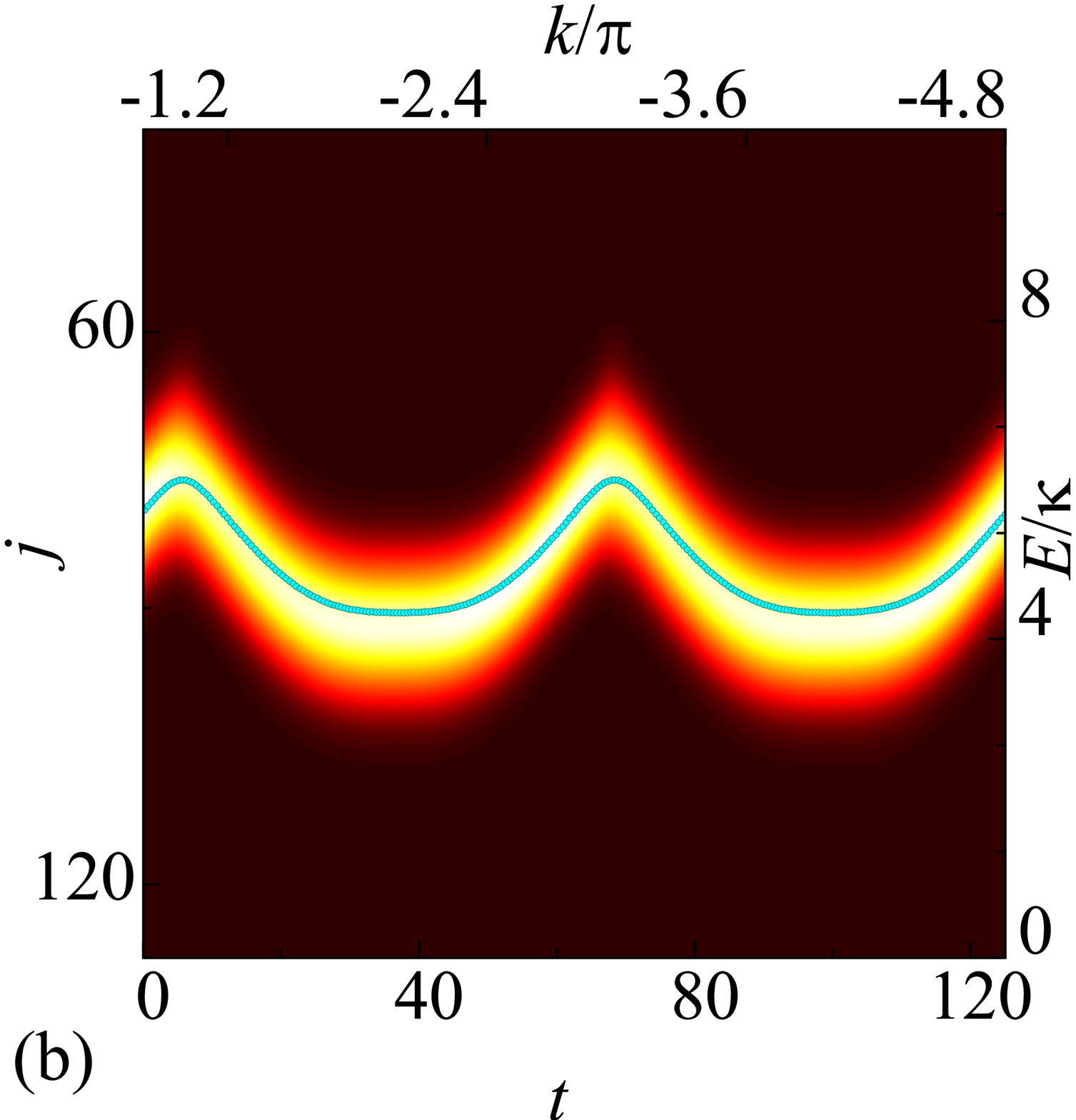} %
\includegraphics[ bb=0 0 594 594, width=0.32\textwidth, clip]{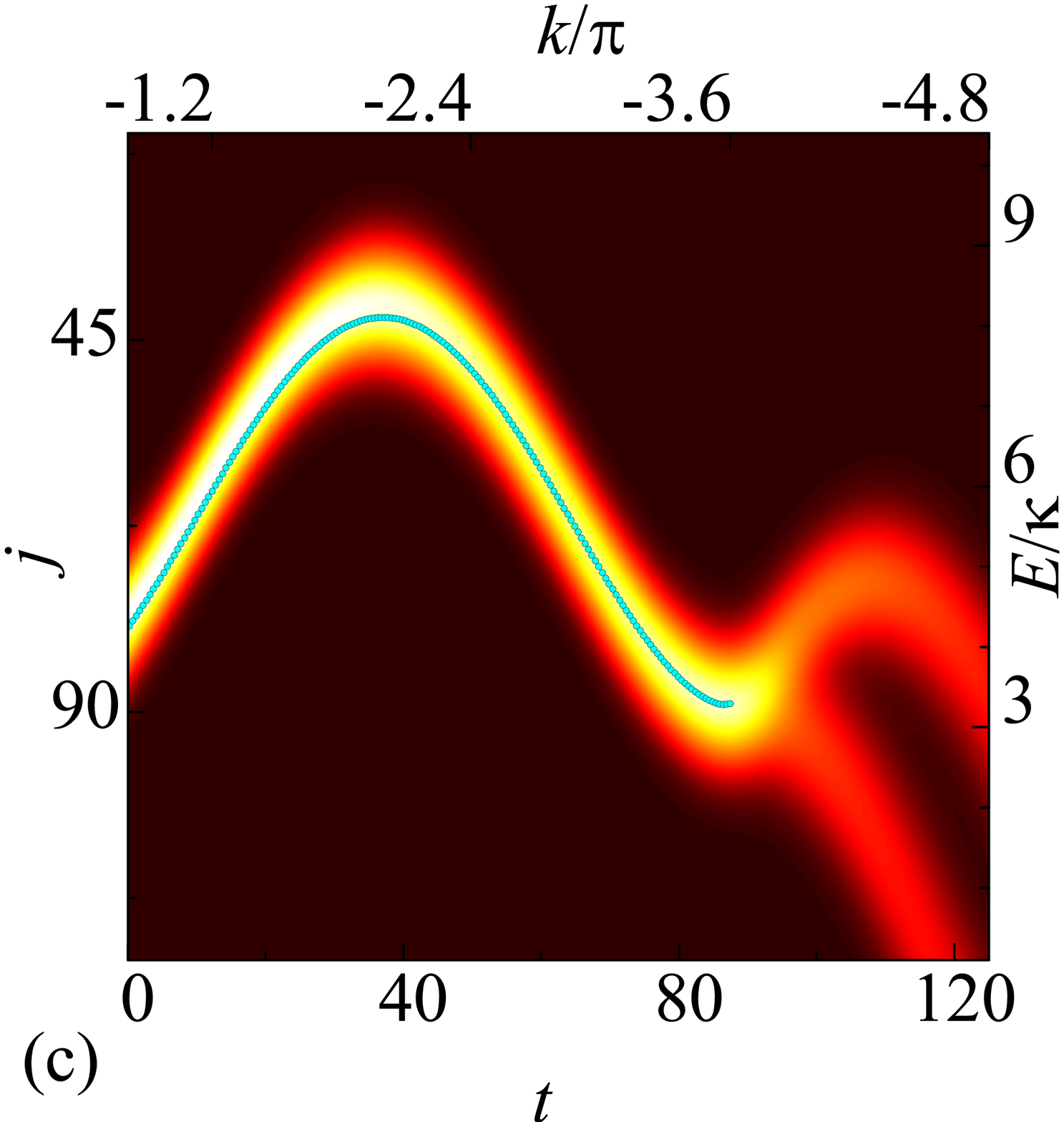}
\par
\caption{(Color online) The comparison between the BP dispersion
relations and the central positions for the cases plotted in Fig. \protect
\ref{fig2} and \protect\ref{fig3}: (a) lower band in $c$; (b) lower band in $%
a$; (c) lower band in $h$. It shows that the two are in close proximity to
each other even for the case of incomplete band, which corresonds to the
sudden death of the BO. The time $t$ is expressed in units of $1/\protect\kappa $. }
\label{fig4}
\end{figure}

Finally, we also plot the BP dispersion relation $E\left( k\right) $
and the central position $x_{\mathrm{c}}\left( t\right) $\ of the wavepacket
under the driving force together in one figure. For several typical cases,
the plots in Fig. \ref{fig4} indicate that the shape of the function $x_{%
\mathrm{c}}\left( t\right) $\ coincides with that of the dispersion relation
$E(k)$. We also find that the semi-classical analysis in Eq. (\ref{CP}) is
valid if $\delta $\ is not too small. Remarkably, one can see that such a
relation still holds even for the incomplete BP band. These results
are in agreement with the theoretical prediction based on the spectral
structures.

\section{Unidirectional propagation}

\label{sec_Unidirectional propagation}

\begin{figure}[tbp]
\centering
\includegraphics[ bb=1 80 581 656, width=0.485\textwidth, clip]{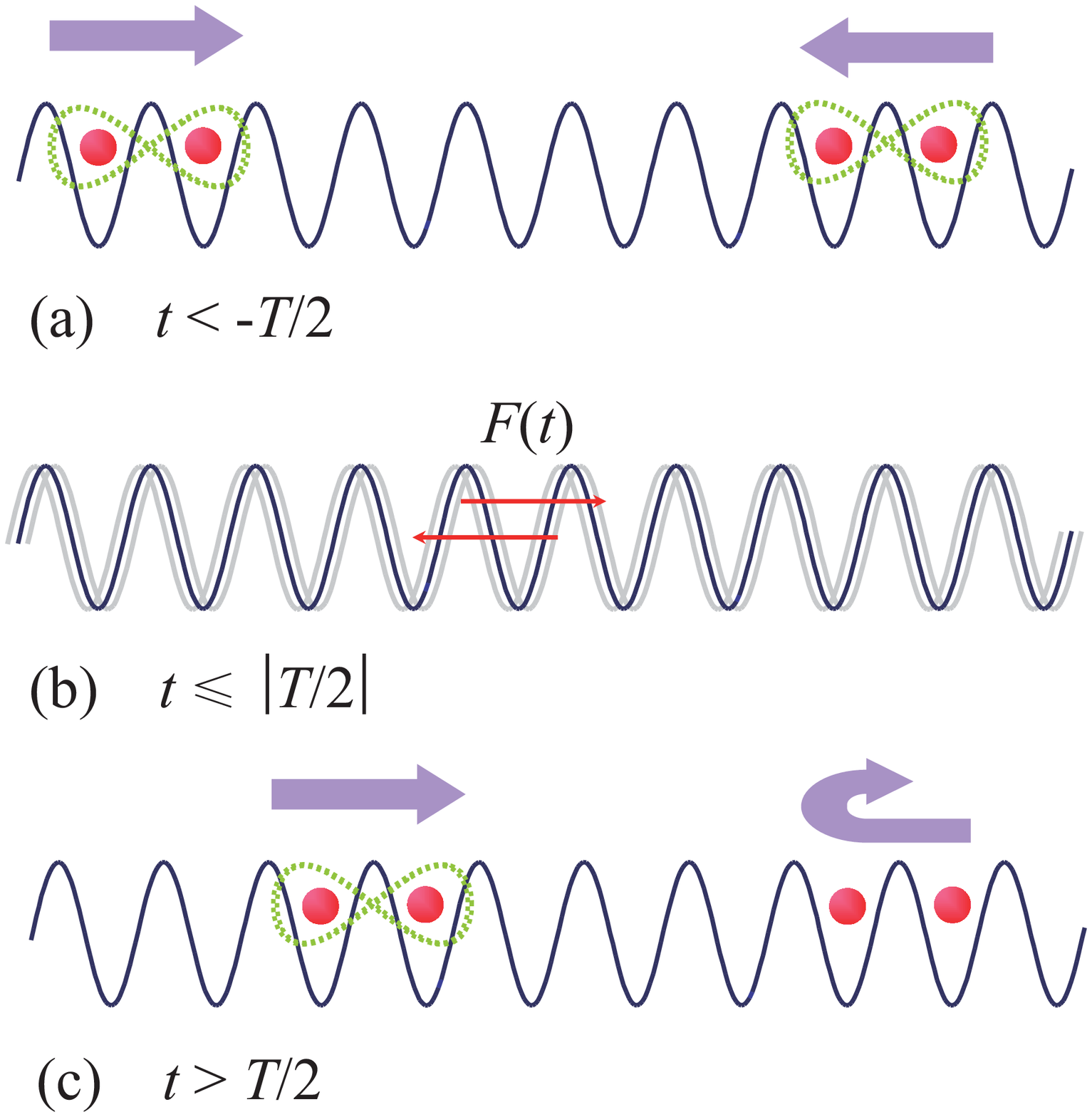} %
\includegraphics[ bb=38 21 573 700, width=0.495\textwidth, clip]{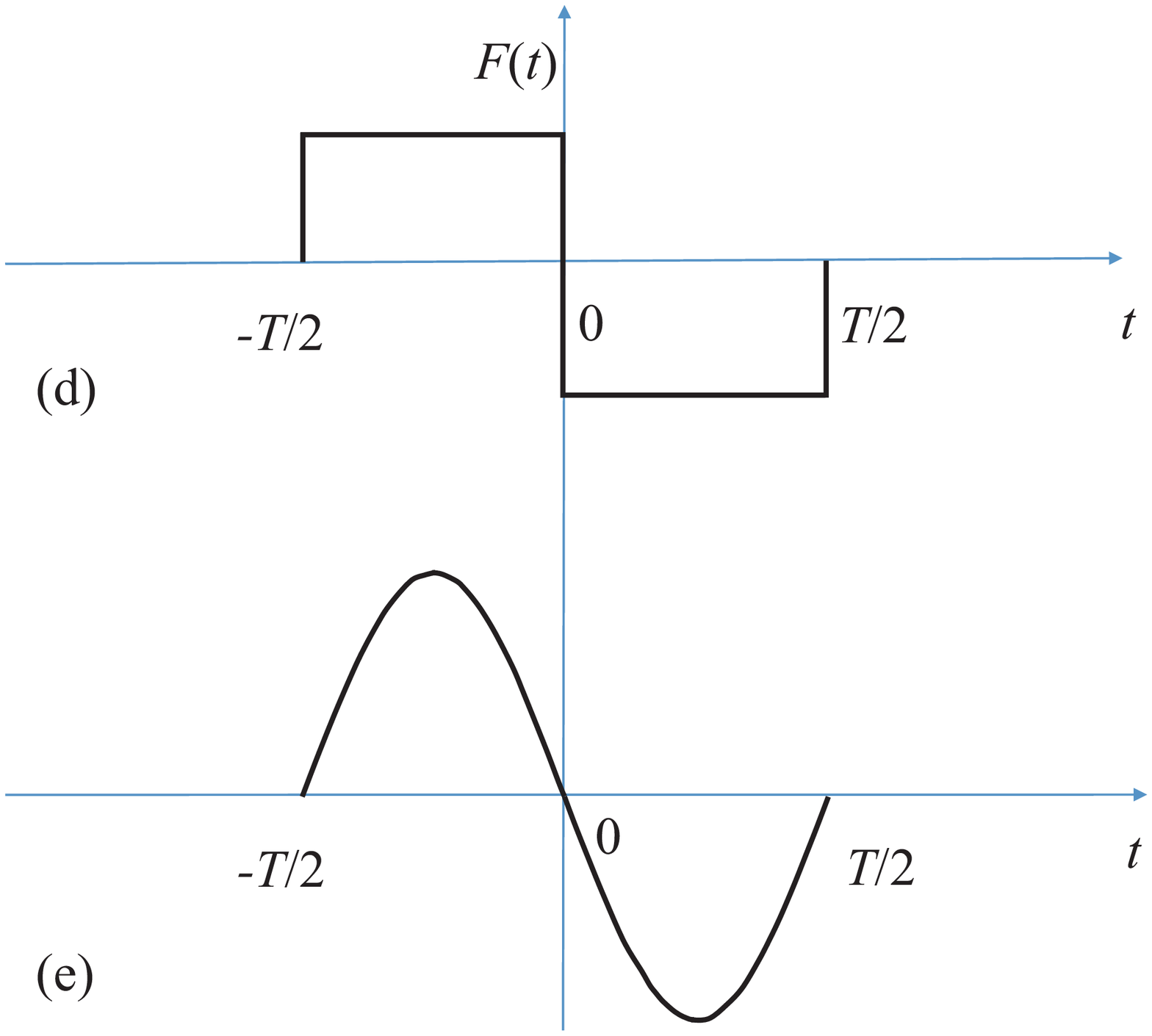}
\par
\caption{(Color online) Schematic illustration for the possess of realizing
unidirectional propagation of BP. The dashed $\infty $\ represents
the correlation between two particles. The shaking lattice induces the
inertial force $F\left( t\right) $. (a) for $t{<-T/2}$, bound-pair
wavepacket moves towards to the right or left with constant group velocity.
(b) during the period of time $[-T/2,T/2]$, a pulse field $F(t)$, is
subjected to the pairs by shaking the lattice back and forth, which may
break down the BP moving to the left. (c) after the time $T/2$, the
surviving wavepacket goes back to its original state, while another pair is
bounded back and becomes uncorrelated. (d) and (e) are two example forms of $%
F(t)$, which are taken for the numerical simulations in Fig. \protect\ref%
{fig6}.}
\label{fig5}
\end{figure}

We now investigate the effect of the time-dependent driving force on the
dynamics of a BP wavepacket. The acceleration theorem (\ref%
{acceleration theorem}) tells us a pulsed field can shift the central
momentum of the wavepacket in the case of complete band. However, it is easy
to find that a pulsed field may destroy a BP wavepacket in the
incomplete band, similarly referred as sudden death of uniform motion. The
death and survival of a propagating wavepacket strongly depends on the
difference between the initial central momentum and the edge of the
incomplete band. Of course, a survived wavepacket can retain its original
motion state\textbf{\ }by a subsequently compensating pulsed field. This
gives rise to a scheme for destroying the pair wavepacket propagating in one
direction, but remaining the one with the opposite direction. Such kind of
scheme can be carried out by two pulsed fields in a pair of adjacent
intervals, which provides two opposite impulses to the wavepacket.

To illustrate the scheme, we propose two concrete examples. The first one is
a square-wave pulse driving force in the form%
\begin{equation}
F\left( t\right) =\left\{
\begin{array}{cc}
F_{0}, & -T/2<t\leq 0, \\
-F_{0}, & 0<t\leq T/2, \\
0, & \textrm{otherwise,}%
\end{array}%
\right. .  \label{Square}
\end{equation}%
According to the acceleration theorem, an initial wavepacket with momentum $%
k_{c}\left( 0\right) $\ will acquire a momentum shift $F_{0}T/2$ at instant $%
t=0$ if $k_{\mathrm{c}}\left( 0\right) +F_{0}T/2$\ is within the band. The
action of the subsequent force $-F_{0}$ can return the group velocity to the
initial value, continue its motion in the same direction. However, in the
case of $k_{\mathrm{c}}\left( 0\right) +F_{0}T/2$\ beyond the incomplete
band, the BP wavepacket breaks down before $t=0$ and the subsequent
force cannot get the correlation back. Therefore, for two BP
wavepackets with opposite momenta $\pm k_{\mathrm{c}}\left( 0\right) $\ or
group velocities $\pm \upsilon _{\mathrm{g}}\left( 0\right) $, one can
always choose a proper $F\left( t\right) $\ to destroy one of them and
maintain the other. This feature can be used to control the direction of
wavepacket propagation\ in demand. Alternatively, one can also consider the
sine-wave pulse driving force

\begin{equation}
F\left( t\right) =\left\{
\begin{array}{cc}
\left( -F_{0}\pi /2\right) \sin \left( 2\pi t/T\right) , & t\leq \left\vert
T/2\right\vert , \\
0, & \textrm{otherwise,}%
\end{array}%
\right. ,  \label{Sine}
\end{equation}%
to achieve the same effect from Eq. (\ref{acceleration theorem}). To examine
how these schemes works in practice, we apply it to a wavepacket in the form
of (\ref{Psi_0}). Fig. 6 shows a numerical propagation of the Gaussian
wavepacket under the action of two kinds of pulse driving forces. It shows
that wavepackets with opposite group velocities exhibit entirely different
behaviors: one remains the original motion state, while the other spreads
out in space.\ Remarkably, the probability of the broken wavepacket is
reflected by the pulsed field, which indicates that the unidirectional
effect in the scheme works not only for the two-particle correlation but
also for the probability flow. In addition, one can see a separation of
slight portion from the moving wavepacket under the action of the
square-wave pulsed field.
\begin{figure}[tbp]
\centering
\includegraphics[ bb=0 0 595 586, width=0.49\textwidth, clip]{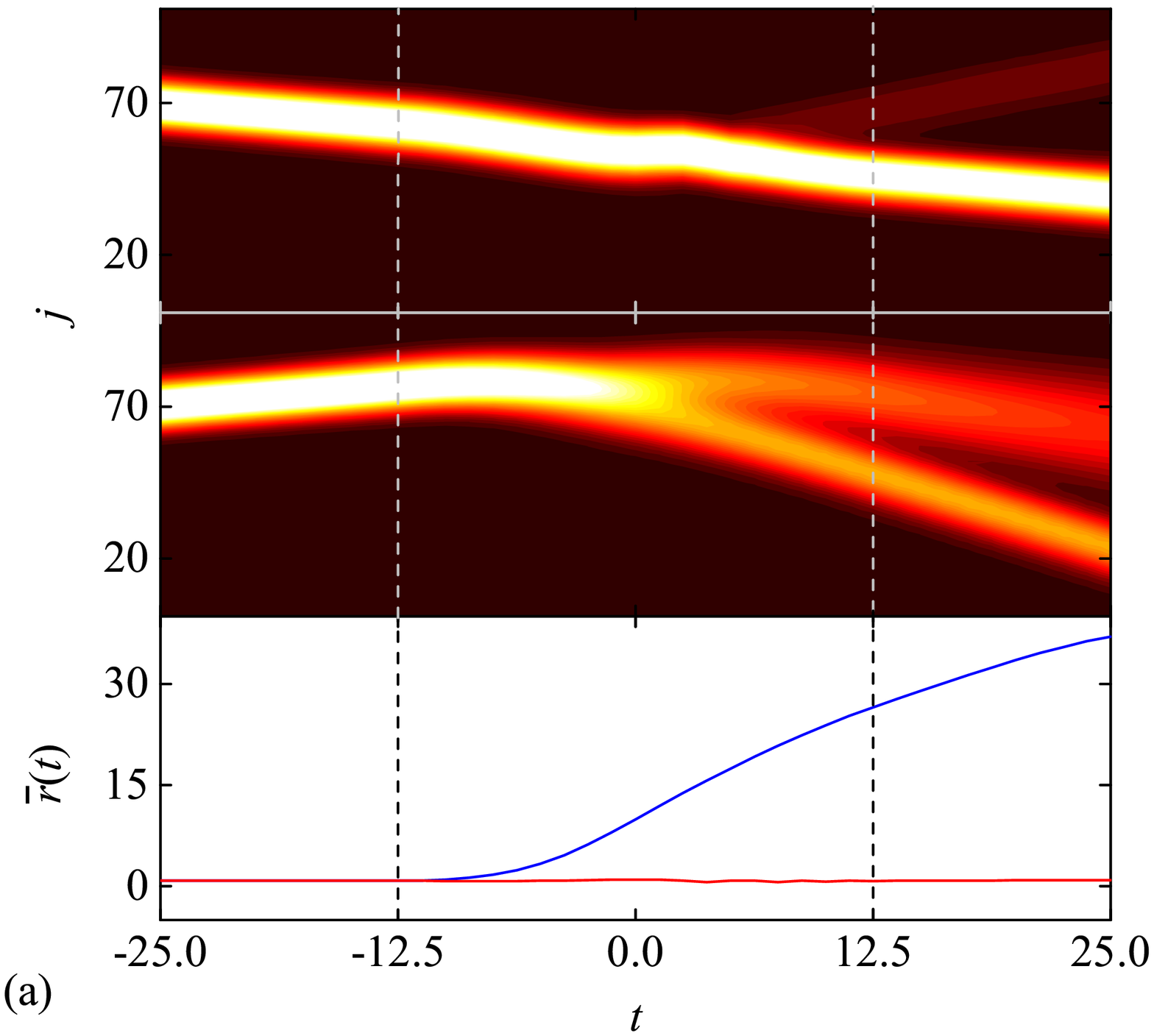} %
\includegraphics[ bb=0 0 595 586, width=0.49\textwidth, clip]{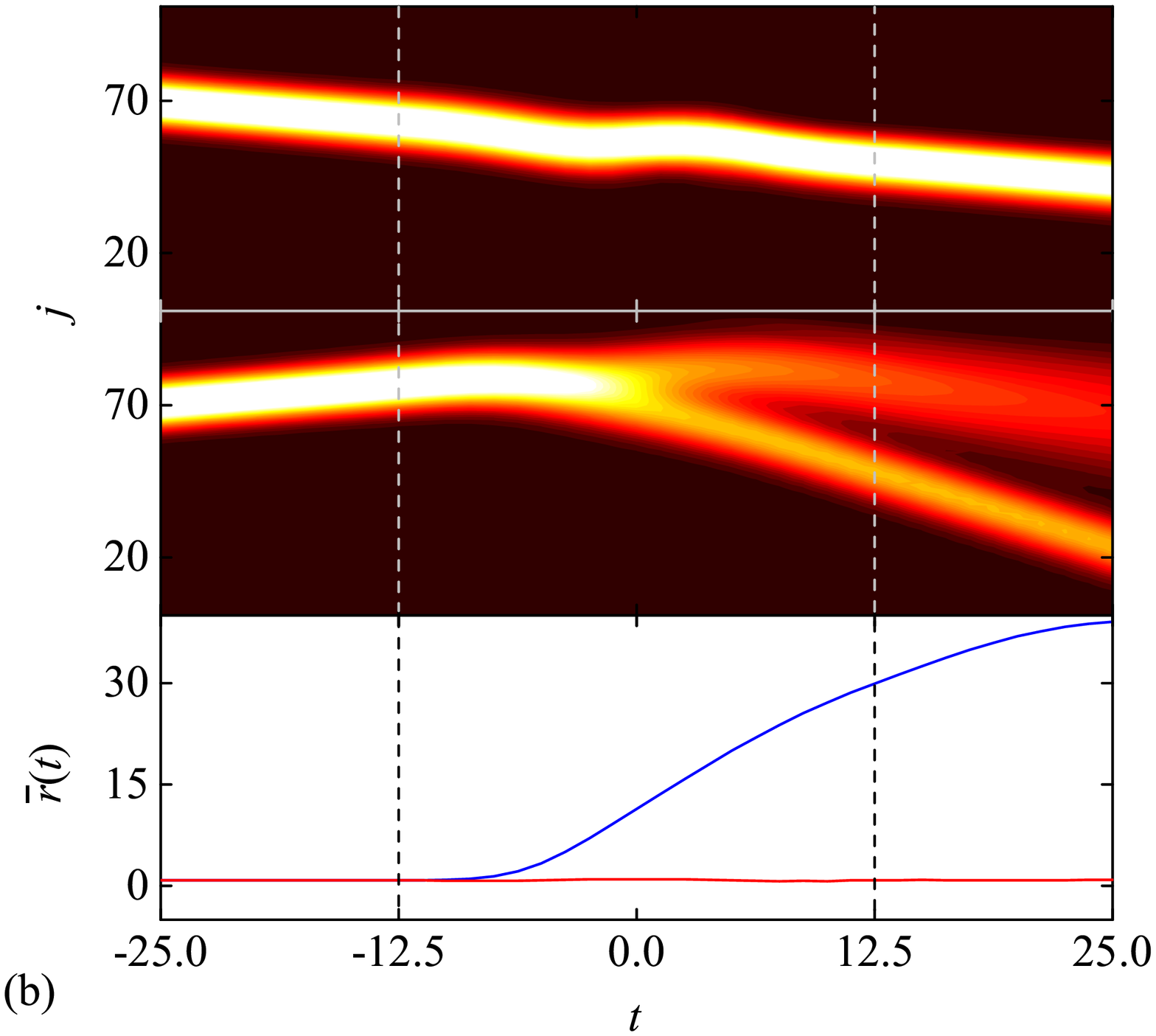}
\par
\caption{(Color online) The profiles and the avarage distances $\bar{r}%
\left( t\right) $\ of the time evolution of the initial wavepackets in the
form of Eq. (\protect\ref{Psi_0}) with $k_{0}=\pm 0.6\protect\pi $, $\protect%
\alpha =0.15$, and $N_{\mathrm{A}}=70,$ in the lower band of the system with
$U=5$, $V=4,$ and $F_{0}=0.05$.\ The pulse field is taken as the forms of
(a) square wave and (b) sine wave, respectively. We can see in both cases
that, the wavepacket with $k_{0}=-0.6\protect\pi $ is spread out by the
pulse field, while the one with $k_{0}=0.6\protect\pi $ remains the original
motion state. Here the time $t$ is expressed in units of $1/\protect\kappa $.}
\label{fig6}
\end{figure}
The probability flow of two particles, no matter correlated or not, can be
depicted by the center of mass (COM)
\begin{equation}
x_{\mathrm{c}}\left( t\right) =\sum_{j}\left\langle jn_{j}\right\rangle _{t},
\label{x_c}
\end{equation}%
where $\left\langle ...\right\rangle _{t}$ denotes the average for a evolved
state. On the other hand, to characterize the efficiency of the schemes, we
introduce the fidelity defined as%
\begin{eqnarray}
\mathcal{F} &=&\textrm{max}\left[ f\left( t\right) \right] ,  \label{f(t)} \\
f\left( t\right)  &=&\left\vert \left\langle \Psi _{t}\left( t\right)
\right\vert \left. \Psi _{0}\left( t_{0}\right) \right\rangle \right\vert ,
\nonumber
\end{eqnarray}%
where%
\begin{equation}
\left\vert \Psi _{0}\left( t_{0}\right) \right\rangle =\exp (-i %
H_{0}t_{0})\left\vert \Psi \left( 0\right) \right\rangle
\end{equation}%
is the target state and
\begin{equation}
\left\vert \Psi _{t}\left( t\right) \right\rangle =\mathcal{T}\exp (-i%
\int_{0}^{t}H\left( t\right) \mathrm{d}t)\left\vert \Psi \left( 0\right)
\right\rangle
\end{equation}%
is the transferred state subjected to a pulsed field. Here we have
omitted a shift on the time scale comparing to the expression of $F\left(
t\right) $. The computation is performed by using a uniform mesh in the time
discretization for the time-dependent Hamiltonian $H\left( t\right) $. As an
example, in Fig. \ref{fig7}, we show the evolution of the COM $x_{\mathrm{c}%
}\left( t\right) \ $and the fidelity $f\left( t\right) $ for the same
parameter values as the four processes simulated in Fig. \ref{fig6}. The
plot in (a) shows the behavior of the two-particle transmission and
reflection induced by the pulsed field, while in (b) the fidelities of the
state transfer. It indicates that a sine-wave pulse has a higher fidelity ($%
\mathcal{F}=0.994$) than that of square-wave pulse ($\mathcal{F}=0.940$).%
\textbf{\ }These results clearly demonstrate the power of the mechanism
proposed in this paper with the purpose to induce unidirectional propagation
which is caused by a pulsed field.
\begin{figure}[tbp]
\centering
\includegraphics[ bb=9 4 501 404, width=0.49\textwidth, clip]{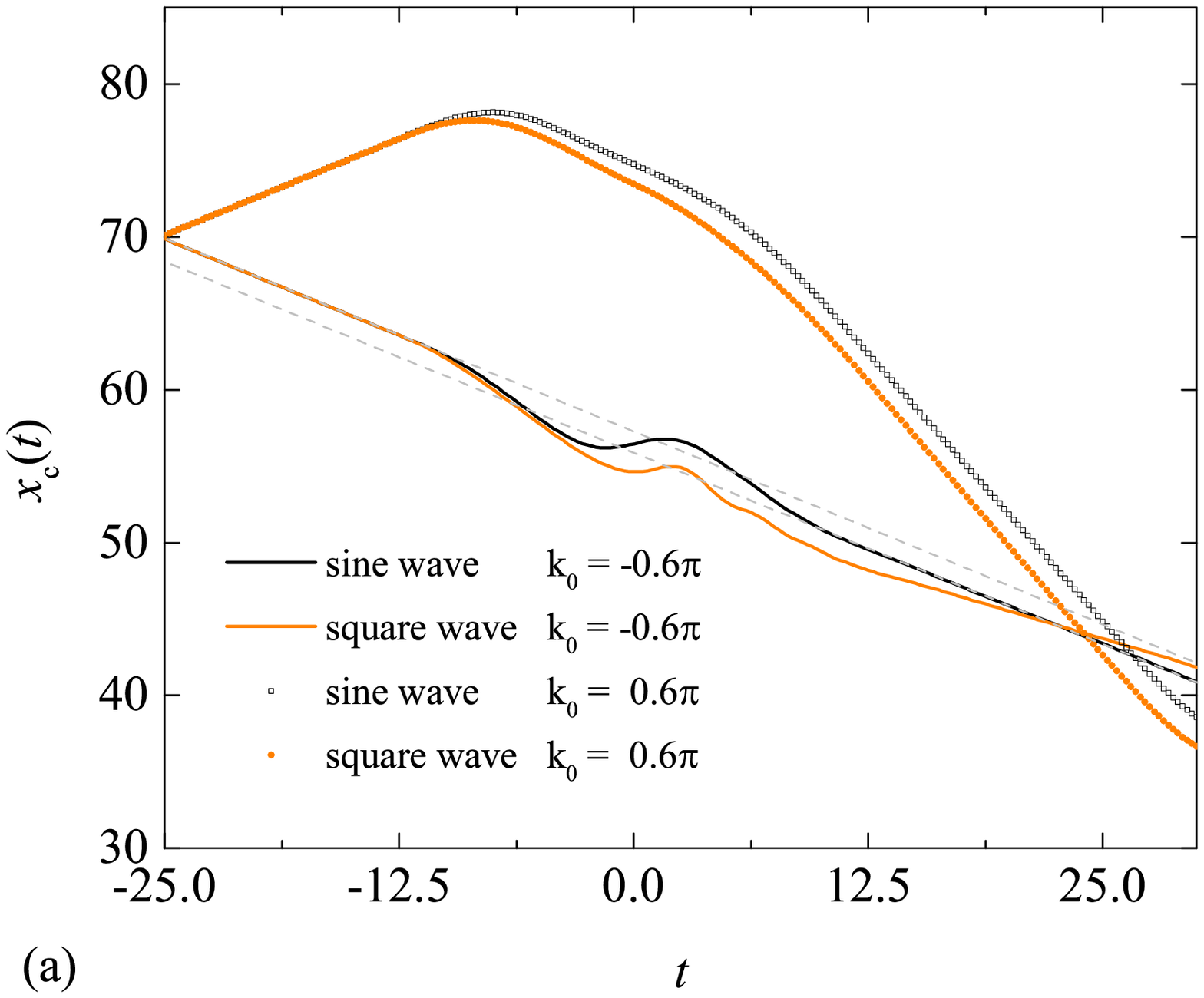} %
\includegraphics[ bb=5 2 488 400, width=0.49\textwidth, clip]{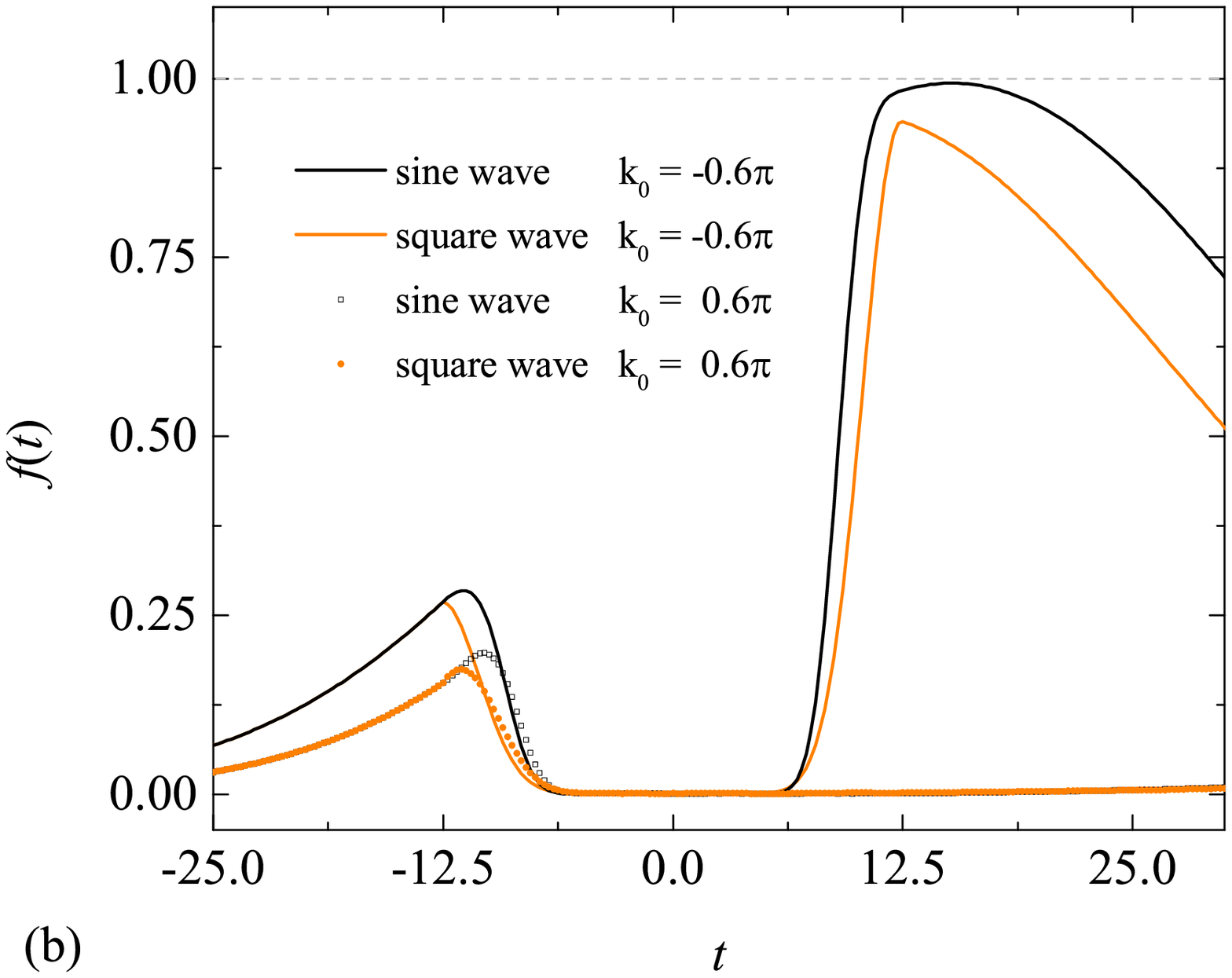}
\par
\caption{(Color online) Plots of the COM $x_{\mathrm{c}}\left( t\right) $\
(a) and the fidelity $f\left( t\right) $\ (b) for the process illustrated in
Fig. \protect\ref{fig6}, where the time $t$ is expressed in units of $1/\protect\kappa $. (a) clearly shows that, the pulsed field bounces back one pair but maitains another. The gray dashed lines are drawn as
guidance to eye indicating the initial and final paths for the case of sine
wave. (b) indicates that the sine-wave pulse has very high efficiency for
the control of unidirectional propagation.}
\label{fig7}
\end{figure}

It is easy to estimate the spectral band of the unidirectional
filter by neglecting the width of the wavepacket. We find that there are
three reasons that trigger the death of a propagating wavepacket: (i) the central
momentum of the initial wavepacket, (ii) the edge of the incomplete band,
which is determined by the\ values of $U/\kappa $ and $V/\kappa $%
, (iii) the impulse of the single pulsed field. We consider the
lattice with one bound band just touching the scattering band at the center
momentum $k=0$. We note, but do not prove exactly, that the
dispersion relation in the left region $\left[ -\pi ,0\right] $
and right region $\left[ 0,\pi \right] $ are monotone functions.
Then if we apply a pulsed field with impulse $\pi $, a moving
wavepacket with the momentum in the left region should be pushed into the
scattering band and will not be recovered by the subsequent $-\pi $ %
impulse. In contrast, a wavepacket in the right region still keep its
initial situation after this process. Therefore, roughly speaking, the
proposed unidirectional filter works for the wavepacket with all possible
group velocity.

\section{Summary}

\label{sec_Summary}

In this paper, the coherent dynamics of two correlated particles in
one-dimensional extended Hubbard model with on-site $U$ and nearest-neighbor
site $V$ interactions, driven by a linear field has been theoretically
investigated. The analysis shows that in the free-field case, there always
exist BP states for any nonzero $U$ and $V$, which may have the
comparable bandwidth with that of single particle. It results in the onset
of distinct BO and BZO for correlated pair in the presence of the external
field. We found that the incompleteness of the BP band spoils the
correlation of the pair and leads to the sudden death of the BO and
BZO.\ Based on this mechanism, we propose a scheme to control the
unidirectional propagation of the BP wavepacket by imposing a single
pulse. As a simple application of this scheme, we investigate the
effect of two kinds of pulsed field. Numerical simulations indicate that a
sine-wave pulse has a higher fidelity than that of square-wave pulse.\ In
experiment, it has been proposed that the ultracold atomic gases in optical
lattices with sinusoidal shaking can be an attractive testing ground to
explore the dynamical control of quantum states \cite%
{Madison,Gemelke,Lignier,Chen}. The sudden death of BO
predicted in this paper is an exclusive signature of correlated particle
pair and could be applied to the quantum and optical device design.

\acknowledgments We acknowledge the support of the National Basic Research Program (973
Program) of China under Grant No. 2012CB921900 and CNSF (Grant No. 11374163).

\end{document}